\documentclass[twocolumn, astrosymb]{aastex701} 

\definecolor{amethyst}{rgb}{0.6, 0.4, 0.8}

\usepackage{graphicx}
\usepackage{amsmath}
\usepackage{amssymb}
\usepackage{makecell}
\usepackage{dirtytalk}
\usepackage{threeparttable}
\usepackage{multirow}
\usepackage{enumitem}
\usepackage{pifont}

\usepackage{soul}

\usepackage{CJKutf8}

\shorttitle{Rest-Frame Optical Nebular Emission Lines at Cosmic Dawn}
\shortauthors{Helton et al.}

\graphicspath{{./}{files/}}

\begin{document}

\newcommand{\ResultBD}{$\mathrm{H} \alpha / \mathrm{H} \beta = 1.7 \pm 0.6$}
\newcommand{\ResultSFR}{$\mathrm{SFR} \approx 8 \pm 2\ M_{\odot} / \mathrm{yr}$}
\newcommand{\ResultSigmaSFR}{$\Sigma_{\mathrm{SFR}} \approx 20 \pm 5\ M_{\odot}/\mathrm{yr}/\mathrm{kpc}^{2}$}
\newcommand{\ResultXiIon}{\mbox{$\xi_{\mathrm{ion}} \approx 10^{25.3 \pm 0.1}\ \mathrm{Hz/erg}$}}
\newcommand{\ResultOtoH}{$\mathrm{log}_{10}(\mathrm{O/H}) + 12 \approx 7.5 \pm 0.2$}
\newcommand{\ResultCtoO}{$[\mathrm{C/O}] \approx -0.4 \pm 0.2$}
\newcommand{\ResultLogU}{$\mathrm{log}_{10}(U) \gtrsim -2.4$}
\newcommand{\ResultDensity}{$n_{\mathrm{H}} \approx 690 \pm 200\ \mathrm{cm}^{-3}$}

\newcommand{\CueResultXiIon}{\mbox{$\xi_{\mathrm{ion}} \approx 10^{25.3 \pm 0.1}\ \mathrm{Hz/erg}$}}
\newcommand{\CueResultOtoH}{$\mathrm{log}_{10}(\mathrm{O/H}) + 12 \approx 8.5_{-0.4}^{+0.4}$}
\newcommand{\CueResultCtoO}{$[\mathrm{C/O}] \approx +0.1_{-0.4}^{+0.4}$}
\newcommand{\CueResultNtoO}{$[\mathrm{N/O}] \approx -0.3_{-0.5}^{+0.6}$}
\newcommand{\CueResultLogU}{$\mathrm{log}_{10}(U) \approx -1.4_{-0.4}^{+0.3}$}
\newcommand{\CueResultDensity}{$n_{\mathrm{H}} \approx 540_{-320}^{+520}\ \mathrm{cm}^{-3}$}

\title{Ionizing Photon Production Efficiencies and Chemical Abundances at Cosmic Dawn \\ Revealed by Ultra-Deep Rest-Frame Optical Spectroscopy of JADES-GS-z14-0}


\suppressAffiliations

\author[0000-0003-4337-6211]{Jakob M. Helton}
\affiliation{Department of Astronomy and Astrophysics, The Pennsylvania State University, University Park, PA 16802, USA}
\email{jakobhelton@psu.edu}

\author[0000-0002-9288-9235]{Jane E. Morrison}
\affiliation{Steward Observatory, University of Arizona, 933 N. Cherry Ave., Tucson, AZ 85721, USA}
\email{janem@arizona.edu}

\author[0000-0003-4565-8239]{Kevin N. Hainline}
\affiliation{Steward Observatory, University of Arizona, 933 N. Cherry Ave., Tucson, AZ 85721, USA}
\email{kevinhainline@arizona.edu}

\author[0000-0003-2388-8172]{Francesco D'Eugenio}
\affiliation{Kavli Institute for Cosmology, University of Cambridge, Madingley Rd., Cambridge CB3 0HA, UK}
\affiliation{Cavendish Laboratory, University of Cambridge, 19 JJ Thomson Ave., Cambridge CB3 0HE, UK}
\email{francesco.deugenio@gmail.com}

\author[0000-0003-2303-6519]{George H. Rieke}
\affiliation{Steward Observatory, University of Arizona, 933 N. Cherry Ave., Tucson, AZ 85721, USA}
\email{ghrieke@gmail.com}


\author[0000-0002-8909-8782]{Stacey Alberts}
\affiliation{AURA for the European Space Agency (ESA), Space Telescope Science Institute, 3700 San Martin Dr., Baltimore, MD 21218, USA}
\email{salberts@stsci.edu}

\author[0000-0002-6719-380X]{Stefano Carniani}
\affiliation{Scuola Normale Superiore, Piazza dei Cavalieri 7, I-56126 Pisa, Italy}
\email{stefano.carniani@sns.it}

\author[0000-0001-6755-1315]{Joel Leja}
\affiliation{Department of Astronomy and Astrophysics, The Pennsylvania State University, University Park, PA 16802, USA}
\affiliation{Institute for Computational and Data Sciences, The Pennsylvania State University, University Park, PA 16802, USA}
\affiliation{Institute for Gravitation and the Cosmos, The Pennsylvania State University, University Park, PA 16802, USA}
\email{joel.leja@psu.edu}

\author[0000-0002-0682-3310]{Yijia Li (\begin{CJK*}{UTF8}{gbsn}李轶佳\end{CJK*})}
\affiliation{Department of Physics and Astronomy, Northwestern University, 2145 Sheridan Rd., Evanston, IL 60208, USA}
\affiliation{Center for Interdisciplinary Exploration and Research in Astrophysics (CIERA), Northwestern University, 1800 Sherman Ave., Evanston, IL 60201, USA}
\email{yijia.li@northwestern.edu}

\author[0000-0002-5104-8245]{Pierluigi Rinaldi}
\affiliation{Space Telescope Science Institute, 3700 San Martin Dr., Baltimore, MD 21218, USA}
\email{prinaldi@stsci.edu}

\author[0000-0001-6010-6809]{Jan Scholtz}
\affiliation{Kavli Institute for Cosmology, University of Cambridge, Madingley Rd., Cambridge CB3 0HA, UK}
\affiliation{Cavendish Laboratory, University of Cambridge, 19 JJ Thomson Ave., Cambridge CB3 0HE, UK}
\email{honzascholtz@gmail.com}

\author[0000-0002-9720-3255]{Meredith Stone}
\affiliation{Steward Observatory, University of Arizona, 933 N. Cherry Ave., Tucson, AZ 85721, USA}
\email{meredithstone@arizona.edu}

\author[0000-0001-9262-9997]{Christopher N. A. Willmer}
\affiliation{Steward Observatory, University of Arizona, 933 N. Cherry Ave., Tucson, AZ 85721, USA}
\email{cnaw@arizona.edu}

\author[0000-0002-8876-5248]{Zihao Wu}
\affiliation{Center for Astrophysics $|$ Harvard and Smithsonian, 60 Garden St., Cambridge, MA 02138, USA}
\email{zihao.wu@cfa.harvard.edu}


\author[0000-0003-0215-1104]{William M. Baker}
\affiliation{DARK, Niels Bohr Institute, University of Copenhagen, Jagtvej 128, DK-2200, Copenhagen, Denmark}
\email{william.baker@nbi.ku.dk}

\author[0000-0002-8651-9879]{Andrew J. Bunker}
\affiliation{Department of Physics, University of Oxford, Denys Wilkinson Building, Keble Rd., Oxford OX1 3RH, UK}
\email{andy.bunker@physics.ox.ac.uk}

\author[0000-0003-3458-2275]{Stephane Charlot}
\affiliation{Sorbonne Universit\'{e}, CNRS, UMR 7095, Institut d'Astrophysique de Paris, 98 bis bd Arago, 75014 Paris, France}
\email{charlot@iap.fr}

\author[0000-0002-7636-0534]{Jacopo Chevallard}
\affiliation{Department of Physics, University of Oxford, Denys Wilkinson Building, Keble Rd., Oxford OX1 3RH, UK}
\email{chevalla@iap.fr}

\author[0000-0001-7151-009X]{Nikko J. Cleri}
\affiliation{Department of Astronomy and Astrophysics, The Pennsylvania State University, University Park, PA 16802, USA}
\affiliation{Institute for Computational and Data Sciences, The Pennsylvania State University, University Park, PA 16802, USA}
\affiliation{Institute for Gravitation and the Cosmos, The Pennsylvania State University, University Park, PA 16802, USA}
\email{cleri@psu.edu}

\author[0000-0002-2678-2560]{Mirko Curti}
\affiliation{European Southern Observatory, Karl-Schwarzschild-Strasse 2, 85748 Garching, Germany }
\email{mirko.curti@eso.org}

\author[0000-0002-9551-0534]{Emma Curtis-Lake}
\affiliation{Centre for Astrophysics Research, Department of Physics, Astronomy and Mathematics, University of Hertfordshire, Hatfield AL10 9AB, UK}
\email{e.curtis-lake@herts.ac.uk}

\author[0000-0003-1344-9475]{Eiichi Egami}
\affiliation{Steward Observatory, University of Arizona, 933 N. Cherry Ave., Tucson, AZ 85721, USA}
\email{egami@arizona.edu}

\author[0000-0002-2929-3121]{Daniel J. Eisenstein}
\affiliation{Center for Astrophysics $|$ Harvard and Smithsonian, 60 Garden St., Cambridge, MA 02138, USA}
\email{deisenstein@cfa.harvard.edu}

\author[0000-0002-6780-2441]{Peter Jakobsen}
\affiliation{Cosmic Dawn Center (DAWN), Copenhagen, Denmark}
\affiliation{Niels Bohr Institute, University of Copenhagen, Jagtvej 128, DK-2200, Copenhagen, Denmark}
\email{pjakobsen@nbi.ku.dk}

\author[0000-0001-7673-2257]{Zhiyuan Ji}
\affiliation{Steward Observatory, University of Arizona, 933 N. Cherry Ave., Tucson, AZ 85721, USA}
\email{zhiyuanji@arizona.edu}

\author[0000-0002-9280-7594]{Benjamin D. Johnson}
\affiliation{Center for Astrophysics $|$ Harvard and Smithsonian, 60 Garden St., Cambridge, MA 02138, USA}
\email{benjamin.johnson@cfa.harvard.edu}

\author[0000-0002-5320-2568]{Nimisha Kumari}
\affiliation{AURA for the European Space Agency (ESA), Space Telescope Science Institute, 3700 San Martin Dr., Baltimore, MD 21218, USA}
\email{kumari@stsci.edu}

\author[0000-0001-6052-4234]{Xiaojing Lin}
\affiliation{Department of Astronomy, Tsinghua University, Beijing 100084, China}
\email{xiaojinglin.astro@gmail.com}

\author[0000-0002-6221-1829]{Jianwei Lyu}
\affiliation{Steward Observatory, University of Arizona, 933 N. Cherry Ave., Tucson, AZ 85721, USA}
\email{jianwei@arizona.edu}

\author[0000-0002-4985-3819]{Roberto Maiolino}
\affiliation{Kavli Institute for Cosmology, University of Cambridge, Madingley Rd., Cambridge CB3 0HA, UK}
\affiliation{Cavendish Laboratory, University of Cambridge, 19 JJ Thomson Ave., Cambridge CB3 0HE, UK}
\affiliation{Department of Physics and Astronomy, University College London, Gower St., London WC1E 6BT, UK}
\email{rm665@cam.ac.uk}

\author[0000-0003-0695-4414]{Michael Maseda}
\affiliation{Department of Astronomy, University of Wisconsin-Madison, 475 N. Charter St., Madison, WI 53706 USA}
\email{maseda@astro.wisc.edu}

\author[0000-0003-4528-5639]{Pablo G. P\'{e}rez-Gonz\'{a}lez}
\affiliation{Centro de Astrobiolog\'{i}a (CAB), CSIC–INTA, Ctra. de Ajalvir km 4, Torrej\'{o}n de Ardoz, E-28850, Madrid, Spain}
\email{pgperez@cab.inta-csic.es}

\author[0000-0002-7893-6170]{Marcia J. Rieke}
\affiliation{Steward Observatory, University of Arizona, 933 N. Cherry Ave., Tucson, AZ 85721, USA}
\email{mrieke@gmail.com}

\author[0000-0002-4271-0364]{Brant Robertson}
\affiliation{Department of Astronomy and Astrophysics, University of California, Santa Cruz, 1156 High St., Santa Cruz, CA 95064, USA}
\email{brant@ucsc.edu}

\author[0000-0001-5333-9970]{Aayush Saxena}
\affiliation{Department of Physics, University of Oxford, Denys Wilkinson Building, Keble Rd., Oxford OX1 3RH, UK}
\affiliation{Department of Physics and Astronomy, University College London, Gower St., London WC1E 6BT, UK}
\email{aayush.saxena@physics.ox.ac.uk}

\author[0000-0002-4622-6617]{Fengwu Sun}
\affiliation{Center for Astrophysics $|$ Harvard and Smithsonian, 60 Garden St., Cambridge, MA 02138, USA}
\email{fengwu.sun@cfa.harvard.edu}

\author[0000-0002-8224-4505]{Sandro Tacchella}
\affiliation{Kavli Institute for Cosmology, University of Cambridge, Madingley Rd., Cambridge CB3 0HA, UK}
\affiliation{Cavendish Laboratory, University of Cambridge, 19 JJ Thomson Ave., Cambridge CB3 0HE, UK}
\email{st578@cam.ac.uk}

\author[0000-0003-4891-0794]{Hannah \"{U}bler}
\affiliation{Max-Planck-Institut f\"{u}r extraterrestrische Physik (MPE), Gie{\ss}enbachstra{\ss}e 1, 85748 Garching, Germany}
\email{hannah@mpe.mpg.de}

\author[0000-0001-8349-3055]{Giacomo Venturi}
\affiliation{Scuola Normale Superiore, Piazza dei Cavalieri 7, I-56126 Pisa, Italy}
\email{giacomo.venturi1@sns.it}

\author[0000-0003-2919-7495]{Christina C. Williams}
\affiliation{NSF National Optical-Infrared Astronomy Research Laboratory, 950 N. Cherry Ave., Tucson, AZ 85719, USA}
\email{christina.williams@noirlab.edu}

\author[0000-0002-4201-7367]{Chris Willott}
\affiliation{NRC Herzberg, 5071 W. Saanich Rd., Victoria, BC V9E 2E7, Canada}
\email{chris.willott@nrc.ca}

\author[0000-0002-7595-121X]{Joris Witstok}
\affiliation{Cosmic Dawn Center (DAWN), Copenhagen, Denmark}
\affiliation{Niels Bohr Institute, University of Copenhagen, Jagtvej 128, DK-2200, Copenhagen, Denmark}
\email{joris.witstok@nbi.ku.dk}

\author[0000-0003-3307-7525]{Yongda Zhu}
\affiliation{Steward Observatory, University of Arizona, 933 N. Cherry Ave., Tucson, AZ 85721, USA}
\email{yongdaz@arizona.edu}


\collaboration{all}{The complete list of authors and affiliations are provided at the end of the manuscript.} 


\correspondingauthor{Jakob M. Helton}
\email{jakobhelton@psu.edu}

\begin{abstract}

JWST has discovered an early period of galaxy formation that was more vigorous than expected, which has challenged our understanding of the early Universe. In this work, we present the longest spectroscopic integration ever acquired by JWST/MIRI ($t_{\mathrm{obs}} \approx 51\ \mathrm{hr}$). This spectrum covers the brightest rest-frame optical nebular emission lines for the luminous galaxy \mbox{JADES-GS-z14-0} at $z = 14.18$. Most notably, we detect $[\mathrm{OIII}] \lambda\lambda 4959{,}5007$ at $\approx 14 \sigma$ and $\mathrm{H}\alpha$ at $\approx 4 \sigma$ with these ultra-deep observations. These lines reveal that JADES-GS-z14-0 has low dust attenuation with a recent star formation rate of \ResultSFR, star formation rate surface density of \ResultSigmaSFR, and ionizing photon production efficiency of \ResultXiIon. Using standard strong-line diagnostics, we infer a gas-phase oxygen abundance of \ResultOtoH\ ($\approx 6\%\ Z_{\odot}$), carbon-to-oxygen ratio of \ResultCtoO, ionization parameter of \ResultLogU, and \mbox{density of \ResultDensity}. Using detailed photoionization modeling, we instead derive \CueResultOtoH\ ($\approx 60\%\ Z_{\odot}$), \CueResultLogU, and \CueResultDensity. The inferred properties of \mbox{JADES-GS-z14-0} are similar to those measured for similarly luminous galaxies at $z > 10$ with previous MIRI/Spectroscopy, such as GHZ2/GLASSz12, GN-z11, and MACS0647-JD1. These results suggest extreme ionization conditions and rapid metal enrichment less than $300$ million years after the Big Bang. Existing simulations are unable to reproduce the empirical and inferred properties of \mbox{JADES-GS-z14-0}. This work demonstrates an important step toward understanding the formation of the first stars and heavy elements in the Universe. Future work will focus on the detection of the rest-frame optical continuum and its interpretation for the stellar population properties of \mbox{JADES-GS-z14-0}. 
 
\end{abstract}


\section{Introduction}
\label{SectionOne}

In its first few years of science operation, JWST has revealed an early period of galaxy formation that was more vigorous than expected. Early results from JWST have shown that a population of luminous galaxies with $M_{\mathrm{UV}} \approx -20$ \citep[e.g.,][]{ArrabalHaro:2023a, ArrabalHaro:2023b, Bunker:2023, Curtis-Lake:2023, Carniani:2024, Castellano:2024, Kokorev:2025, Naidu:2026} and supermassive black holes \citep[e.g.,][]{Goulding:2023, Harikane:2023, Kocevski:2023, Larson:2023, Greene:2024, Maiolino:2024a, Maiolino:2024b, Matthee:2024} already existed less than a billion years after the Big Bang. The observed number densities for these sources are discrepant with predictions from theoretical models and cosmological simulations, sparking debate about whether our understanding of the early Universe, and possibly even the standard cosmological model, needs revision \citep[e.g.,][]{Boylan-Kolchin:2023}.

The most notable and luminous example for this early period of galaxy formation at $z > 12$ is \mbox{JADES-GS-z14-0} at $z = 14.18$ \citep[][]{Carniani:2024, Carniani:2025, Schouws:2025a}. It is currently one of only two galaxies that have been spectroscopically confirmed at $z > 14$, alongside \mbox{MoM-z14} at $z = 14.44$ \citep[][]{Naidu:2026}. \mbox{Beyond} the extreme redshift of \mbox{JADES-GS-z14-0}, there are many other remarkable properties of this galaxy. First, it is spatially resolved with a half-light radius of $r_{\mathrm{UV}} = 260 \pm 20\ \mathrm{pc}$ \citep[][]{Carniani:2024}, which implies the emission at these wavelengths is dominated by stars and nebular gas rather than by an active galactic nucleus (AGN). This galaxy is exceptionally luminous with an absolute UV magnitude of $M_{\mathrm{UV}} = -20.81 \pm 0.16$ \citep[][]{Carniani:2024}, a clear confirmation of the slow decline in the number density of luminous galaxies at $z > 12$ \citep[][]{Robertson:2024, Whitler:2025, Weibel:2026}, contrasting with predictions prior to the launch of JWST. \mbox{JADES-GS-z14-0's} rest-frame UV continuum slope $\beta_{\mathrm{UV}} = -2.20 \pm 0.07$ \mbox{\citep[][]{Carniani:2024}} shows that it is redder than expected for a galaxy at this redshift; its redness is likely caused by evolved stellar populations, significant dust attenuation, relatively high gas-phase metallicities, or enhanced nebular continuum emission. Finally, perhaps the most intriguing \mbox{property} of \mbox{JADES-GS-z14-0} is the significant ($\mathrm{S/N} \approx 13$) photometric detection at $7.7\ \mu\mathrm{m}$ with JWST's Mid-Infrared Instrument \citep[MIRI;][]{Helton:2025}. The observation with JWST/MIRI suggests that \mbox{JADES-GS-z14-0} is metal enriched, which has been confirmed by detections of the emission lines $\mathrm{CIII]} \lambda\lambda 1907{,}1909$ \citep[][]{Carniani:2024} using the JWST Near-Infrared Spectrograph (NIRSpec) along with $\mathrm{[OIII]} \lambda 88 \mu\mathrm{m}$ using the Atacama Large Millimeter/submillimeter Array \citep[ALMA;][]{Carniani:2025, Schouws:2025a}. A second set of observations with ALMA targeted $\mathrm{[CII]} \lambda 158 \mu\mathrm{m}$, but the line was not detected \citep[][]{Schouws:2025b}, although this does not necessarily mean the galaxy is gas poor; JADES-GS-z14-0 is likely embedded in a substantial, pristine reservoir of neutral gas that dominates its total baryon content \citep[][]{Heintz:2025}. Altogether, these results suggest rapid mass assembly and metal enrichment during the earliest phases of galaxy formation, less than $300$ million years after \mbox{the Big Bang}.

Nearly $10$ days of JWST mission time have already been invested in observing JADES-GS-z14-0 \citep[][]{Carniani:2024, Robertson:2024, Helton:2025}. However, the existing observations provide an ambiguous picture of this distant galaxy because of degeneracies in the inferred physical properties. For example, the $R_{3}$ index (defined as $[\mathrm{OIII}] \lambda 5007 / \mathrm{H}\beta$) is often used to probe a galaxy's gas-phase metallicity since this measurement is strongly correlated with the gas-phase oxygen abundance $\mathrm{O/H}$. Using photometry from JWST/MIRI and JWST/NIRCam, \citet[][]{Helton:2025} indirectly derived $R_{3} \approx 2.5$, suggesting metallicities that are less than $10\%$ solar. Using the same set of photometry, \citet[][]{Ferrara:2024} indirectly derived $R_{3} \approx 0.5$, suggesting metallicities that are roughly $1\%$ solar. Finally, using those observations along with spectroscopy from ALMA and JWST/NIRSpec, \citet[][]{Carniani:2025} indirectly derived $R_{3} \approx 5.6$, suggesting metallicities that are roughly $20\%$ solar. Directly measuring the strengths of the rest-frame optical emission lines for \mbox{JADES-GS-z14-0} would provide a robust measurement of the gas-phase metallicity and allow for powerful constraints on models for this galaxy. JWST/MIRI is the only instrument capable of obtaining a rest-frame optical spectrum for this galaxy and others at the redshift frontier (i.e., $z > 10$).

Acquiring spectroscopy with JWST at both rest-frame UV and optical wavelengths is critical for simultaneously understanding the properties of the stellar populations, nebular gas, and dust in the very earliest galaxies \citep[e.g., see][]{Calabro:2024}. Excitingly, MIRI/Spectroscopy has already revealed important insights into galaxies at the redshift frontier \citep[e.g.,][]{Hsiao:2024a, Alvarez-Marquez:2025, Zavala:2025}. Even beyond the gas-phase metallicity, the observations of rest-frame optical emission lines have provided a wealth of information about the properties of these galaxies, including the density, excitation, and ionization states of the interstellar medium in addition to the recent star formation rate and ionizing photon production efficiency of the stellar populations \citep[for a review of understanding galaxy evolution through emission lines, see][]{Kewley:2019}.

In this work, we provide a first look at ultra-deep follow-up observations of \mbox{JADES-GS-z14-0} with MIRI's Low-Resolution Spectrometer (LRS); this is the longest spectroscopic integration ever acquired by JWST/MIRI. Our paper proceeds as follows. In Section~\ref{SectionTwo}, we \mbox{describe} the observations used in our analysis. In Section~\ref{SectionThree}, we explain the process for data reduction, including manual postprocessing and custom corrections that are specialized for long MIRI/LRS integrations. In Section~\ref{SectionFour}, we describe the emission line measurements. In Section~\ref{SectionFive}, we present the physical properties of \mbox{JADES-GS-z14-0} as inferred from strong-line diagnostics using some of the brightest rest-frame optical emission lines along with the previously measured UV and far-infrared lines. This includes diffuse dust attenuation (Section~\ref{SectionFiveOne}), recent star formation rate (Section~\ref{SectionFiveTwo}), ionizing photon production efficiency (Section~\ref{SectionFiveThree}), gas-phase oxygen abundance (Section~\ref{SectionFiveFour}), ionization parameter (Section~\ref{SectionFiveFive}), and electron density (Section~\ref{SectionFiveSix}). In Section~\ref{SectionSix}, we again present the physical properties of \mbox{JADES-GS-z14-0}, but inferred using photoionization modeling of those same lines. In Section~\ref{SectionSeven}, we discuss the interpretation of the inferred physical properties by placing these in the more general context of galaxy formation and evolution in the early Universe. Finally, in Section~\ref{SectionEight}, we summarize our results and their broader implications for understanding galaxies at the break of cosmic dawn. Throughout this work, magnitudes are provided in the AB system \citep[][]{Oke:1983} while uncertainties are quoted as $68\%$ confidence intervals, unless otherwise stated. We report wavelengths in air and assume the standard flat $\Lambda$CDM cosmology from Planck18 with $H_{0} = 67.4\ \mathrm{km/s/Mpc}$ and $\Omega_{m} = 0.315$ \citep[][]{Planck:2020}. We adopt solar abundances as $12 + \mathrm{log}_{10}(\mathrm{O/H}) = 8.69$ and $\mathrm{log}_{10}(\mathrm{C/O}) = -0.23$ \citep[e.g.,][]{Asplund:2021}.

\section{Observations}
\label{SectionTwo}

\begin{figure}
    \centering
    \includegraphics[width=1.0\linewidth]{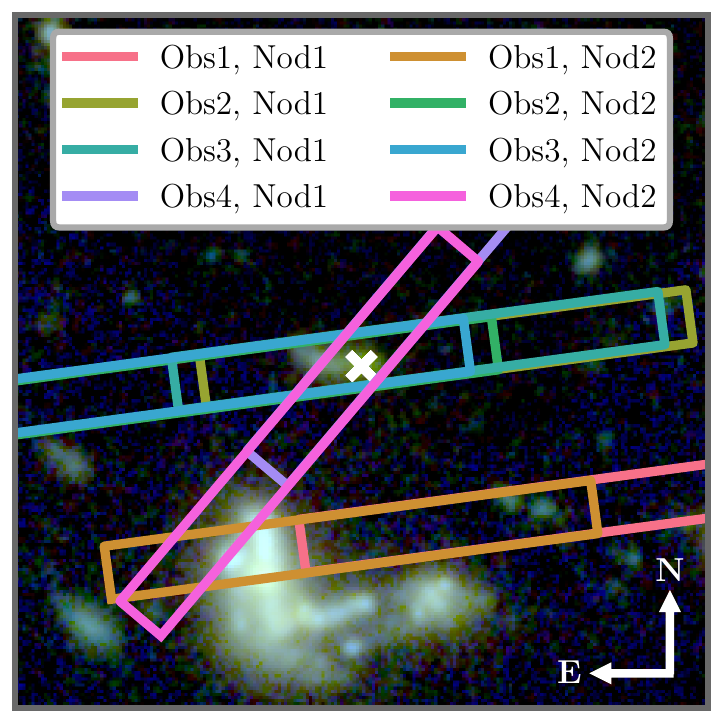}
    \caption{\textbf{Slit locations for four separate visits.} Our target, JADES-GS-z14-0, was observed using the MIRI/LRS in slit mode across four separate visits with two dithers along the slit nod for each visit. We show the eight distinct slit locations alongside the JWST/NIRCam imaging (using F444W-F277W-F115W filters as an RGB false-color mosaic). MIRI/LRS's slit size is $4.7\ \mathrm{arcsec}$ in length and $0.51\ \mathrm{arcsec}$ in width. Three-fourths of the observations were \mbox{successful} in placing JADES-GS-z14-0 within the slit (observations $\#2$, $\#3$, and $\#4$); the remaining one-fourth of the observations were not successful because of failed target acquisition caused by an anomalous cosmic-ray event (observation $\#1$). \label{fig:Slit_Locations}}
\end{figure}

The primary suite of observations presented in this work includes ultra-deep, low-resolution, mid-infrared \mbox{spectroscopy} of \mbox{JADES-GS-z14-0} from JWST/MIRI \citep[][]{Rieke:2015, Wright:2023}. The first three sets of observations were obtained by JWST program ID GO/8544 (PI: J.~Helton) on $15\mathrm{-}22$ November $2025$ using the MIRI/LRS \citep[][]{Kendrew:2015, Kendrew:2016} in slit mode for a total on-source integration time of $183.8\ \mathrm{ks}$ ($51.0\ \mathrm{hr}$). The first two-thirds of the observations were successful in placing JADES-GS-z14-0 within the slit (observations $\#2$ and $\#3$); however, the final one-third of observations were unsuccessful in slit placement due to failed target acquisition (TA) caused by an anomalous cosmic-ray event (observation $\#1$). Because of this failure, a fourth set of observations was obtained on $1\mathrm{-}2$ January $2026$ for a total on-source integration time of $61.3\ \mathrm{ks}$ ($17.0\ \mathrm{hr}$). Figure~\ref{fig:Slit_Locations} illustrates the on-sky slit locations from these four sets of observations. In this work, we only use the successful three-fourths of the acquired data for a total integration time of $183.8\ \mathrm{ks}$ ($51.0\ \mathrm{hr}$). Our observing strategy was similar to that of program ID GO/3703 (PI: J.~Zavala), where the luminous galaxy \mbox{GHZ2/GLASSz12} at $z = 12.34$ was observed using the MIRI/LRS for a total of $32.5\ \mathrm{ks}$ \citep[$9.0\ \mathrm{hr}$;][]{Zavala:2025}. However, there are several notable differences in our observing strategy,  described in detail below.

Our target, JADES-GS-z14-0, was observed by four separate visits using the P750L filter and the FASTR1 readout pattern with $119$ groups per integration, $23$ \mbox{integrations} per exposure, four exposures per dither, and two dithers along the slit nod. The four exposures per dither were obtained using a two-by-two mosaic with $100\%$ overlap for the rows and columns. We chose to have multiple visits in order to provide additional \mbox{dithering} between both of the slit locations. All four visits were offset from the centroid of \mbox{JADES-GS-z14-0} by one pixel (i.e., $0.11\ \mathrm{arcsec}$) in the $x$-direction to avoid previously known bad pixels overlapping with the locations of the most prominent rest-frame optical emission lines. The second and third visits were \mbox{additionally} offset from the first and fourth visits by two pixels (i.e., $0.22\ \mathrm{arcsec}$) in either \mbox{direction}. The first three visits were executed at a position angle (V3PA) of $\approx 3^{\circ}$ for a total of $24$ exposures across all three visits. The fourth visit was executed at a position angle (V3PA) of $\approx 45^{\circ}$ for a total of eight exposures. As a reminder, we only use three-fourths of the data for a total of $24$ exposures.

For each of the visits, direct images were acquired with the MIRI/F560W filter for initial TA and subsequent verification after repositioning on \mbox{JADES-GS-z14-0}. We obtained verification images because they are strongly recommended when obtaining slit spectroscopy with the MIRI/LRS to confirm accurate slit placement. We chose a relatively bright ($m \approx 21\ \mathrm{AB\ mag}$ in MIRI/F560W) nearby star located roughly $40\ \mathrm{arcsec}$ to the northeast of \mbox{JADES-GS-z14-0} as the target for TA. For TA, we used the \mbox{FASTGRPAVG} readout pattern with $10$ groups per integration and one integration per exposure, while for verification, we used the FASTR1 readout pattern with $119$ groups per integration and one integration per exposure. We used the verification images to confirm that TA was successful for three out of the four visits. The failed TA for the remaining visit centered on a bright, unflagged cosmic-ray event instead of centroiding on the bright nearby star that we chose. The failed TA was possibly caused by elevated solar activity immediately before and throughout the observing window, which ultimately led to an increased number of cosmic-ray events affecting our observations, both in the imaging and spectroscopic exposures, as described in more detail below.

\section{Data Reduction}
\label{SectionThree}

To reduce the data obtained by our set of MIRI/LRS observations, we used the most recently updated version of the standard JWST Calibration Pipeline (version $2.0.0$) and Calibration Reference Data System (version $13.1.14$) pipeline mapping ($1536$). This multistage process transforms the raw data obtained by JWST into scientifically viable spectra by systematically correcting for various instrumental artifacts. Stage~1 of the calibration pipeline involves bad pixel flagging, dark current subtraction, cosmic-ray shower removal, nonlinearity corrections, adjustments for reset effects, gain calibration, and ramp fitting. Stage~2 of the pipeline involves setting up the world coordinate system (WCS), flat-field corrections, flux calibration, background subtraction, wavelength calibration, and initial spectral extraction. Stage~3 of the pipeline involves outlier detection, image combination, error propagation, and final spectral extraction. While STScI's standard pipeline addresses the most common requirements for calibration and reduction, the extreme depth and complexity of our observations presented unique challenges that required manual postprocessing and customized corrections to obtain the data quality needed for effective scientific interpretation of the data. The modifications that we developed for the pipeline are described in detail below.

\begin{figure*}
    \centering
    \includegraphics[width=0.9\linewidth]{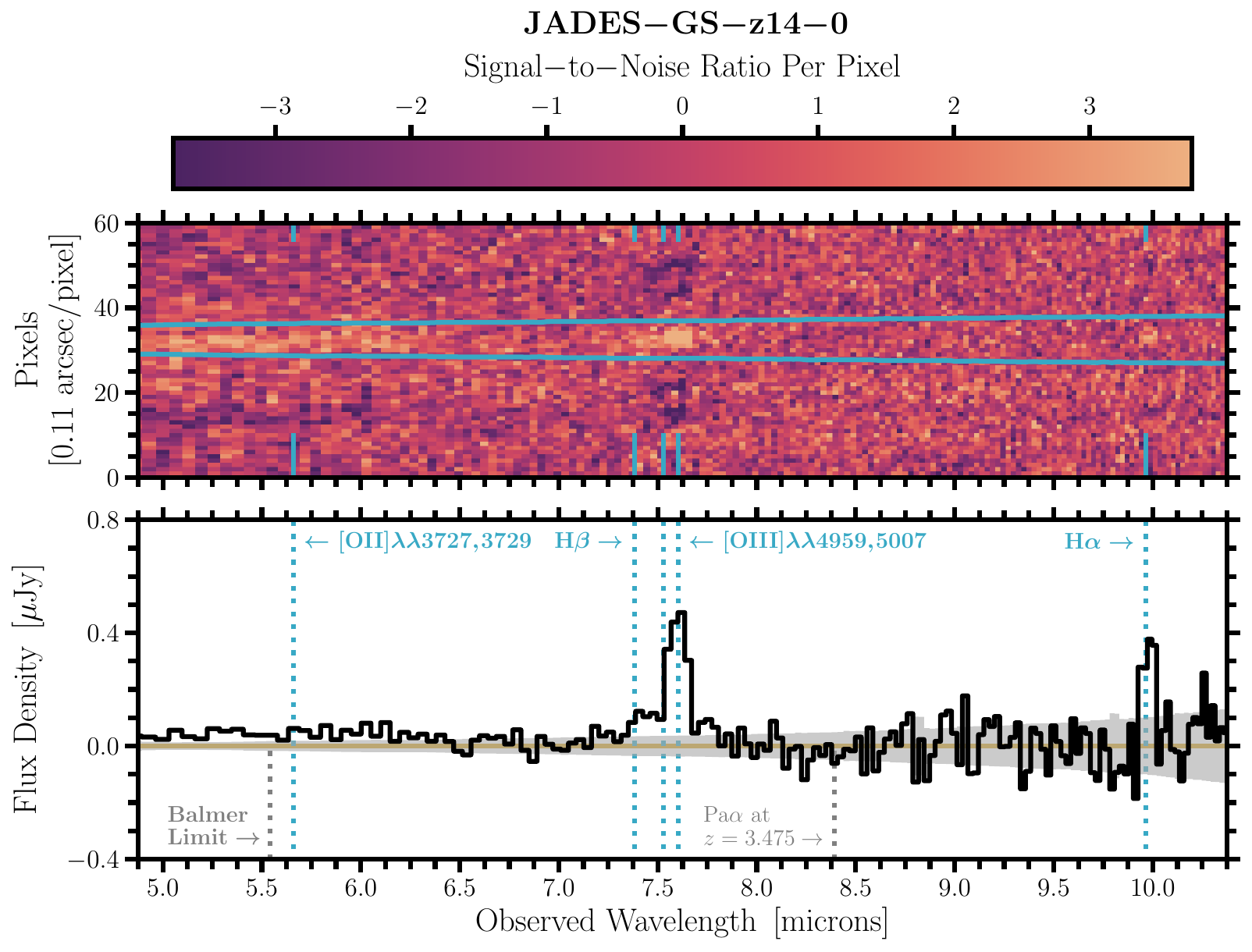}
    \caption{\textbf{The rest-frame optical spectrum of JADES-GS-z14-0.} \textit{Top panel:} The final MIRI/LRS 2D spectrum is provided. The corresponding color bar for the measured signal-to-noise ratio per pixel is also shown. JWST/MIRI's detector plate scale is $0.11\ \mathrm{arcsec/pixel}$, as shown by the $y$-axis label. Blue, horizontal lines roughly indicate the wavelength-dependent profiles used for the 1D optimal extraction while the blue, vertical lines indicate the locations for some of the strongest rest-frame optical emission lines. \textit{Bottom panel:} The final MIRI/LRS 1D spectrum is provided. There are two spectroscopic features that are clearly detected above the noise level ($> 4 \sigma$) in both the 2D and 1D spectra, in addition to the continuum, which is marginally detected ($> 1 \sigma$ per wavelength bin at $\lambda_{\mathrm{obs}} \lesssim 6\ \mu\mathrm{m}$). The detection of these rest-frame optical emission lines and their scientific interpretation are the focus of this work, while the rest-frame optical continuum will be discussed and interpreted in a forthcoming manuscript (Helton et~al. 2026, \mbox{in preparation}). The gray, vertical lines indicate the locations of the Balmer continuum limit for JADES-GS-z14-0 and the $\mathrm{Pa}\alpha$ line for the neighboring foreground galaxy at $z = 3.475$ (see also Figure~\ref{fig:Slit_Locations}). \label{fig:Full_Spectrum}}
\end{figure*}

Initially, we ran the pipeline with the default configuration, but ultimately made the following revisions to the pipeline's default parameters. We first run Stage~1 of the pipeline with ``find\_showers'' set to true and ``only\_use\_ints'' set to false for the jump detection step in order to better deal with pixels affected by cosmic-ray events. Immediately after Stage~1, we correct the telescope's (V2, V3) reference pixels in the headers of the rate files to account for an additional offset due to the initial pointing inaccuracy. We calculate this offset for each visit separately by using the verification images for source identification and comparing with the brightest objects in the JADES photometric catalogs. In units of native pixels from JWST/MIRI, and assuming a detector plate scale of $0.11\ \mathrm{arcsec/pixel}$, we derive (V2, V3) offsets of ($0.579$, $1.616$), ($0.761$, $1.669$), ($0.335$, $1.883$), and ($0.213$, $1.497$) for observations $\#1$, $\#2$, $\#3$, and $\#4$, \mbox{respectively}. We additionally clean the rate files by individually sigma clipping each pixel. For each nodded dither of each visit, we mask sigma-clipped outliers across the $92$ integrations ($23$ integrations per exposure and four exposures per dither, so $23 \times 4$) by identifying pixels deviating from the median value by more than $3 \sigma$, where $\sigma$ refers to the standard deviation derived from the median absolute deviation; this metric is less sensitive to outliers than a typical standard deviation.

Because we noticed persistent cosmic-ray events in the rate files, even after masking sigma-clipped outliers, we also mask five rows along the trace for observation $\#3$ and three rows along the trace for observation $\#4$. For reference, rows are arranged along the cross-dispersion direction (i.e., the spatial direction) whereas columns are arranged along the dispersion direction (i.e., the wavelength direction). We additionally mask three columns in the second nod of observation $\#4$ because of significant contamination from a foreground galaxy at a separation of roughly $2.2\ \mathrm{arcsec}$ to the south (with NIRCam ID 182698 and a spectroscopic redshift of $z = 2.032$ from emission lines; see also Figure~\ref{fig:Slit_Locations}). We experimented with using a master background technique (as commonly used to reduce JWST/MIRI data) to avoid masking so much data when dealing with the contaminating galaxy, but found this technique to be inadequate for our purposes because it noticeably reduced the final data quality.

We then run Stage~2 of the pipeline with nodded background subtraction and path loss corrections included. For path loss corrections, we assume \mbox{JADES-GS-z14-0} is a point source (``source\_type'' set to ``POINT'') since JWST/MIRI's point-spread function (PSF) is notably larger than JADES-GS-z14-0 (with half-light radius of $r_{\mathrm{UV}} = 0.079 \pm 0.006\ \mathrm{arcsec}$) across the complete spectral range of the MIRI/LRS (full-width at half-maximum of $\mathrm{FWHM} > 0.2\ \mathrm{arcsec}$ at $\lambda_{\mathrm{obs}} > 5.0\ \mu\mathrm{m}$). In Stage~2, we also specify offsets for each visit (see the discussion in Section~\ref{SectionTwo} for more information about this) while saving results for both the nodded background subtraction and assign WCS steps; intermediate results, such as these ones, are visually inspected at every stage of the data reduction process to validate the various data products. Upon visual inspection, we noticed residual backgrounds in many of the cal files. One of the most prominent residuals was the ``bar'' artifact, which is effectively a persistence imprint of the MIRI/LRS slit on the spectrum at $\lambda_{\mathrm{obs}} \approx 8.0-8.5\ \mu\mathrm{m}$. Any exposure that begins with TA will have this feature, which quickly decays and typically disappears by the third exposure following TA. To address this feature in our data, and any other issues with the global background, we individually subtract median values from each row of each exposure.

Lastly, we run Stage~3 of the pipeline and perform optimal extraction (``extraction\_type'' set to ``optimal'') with ``model\_nod\_pair'' and ``use\_source\_posn'' both set to true, but ``subtract\_background'' set to false since we already included nodded background subtraction in Stage~2. We do not include the pixel replacement step since we do not want flagged pixels to be included in the optimal extraction. The association files for Stage~3 were modified to include all $24$ exposures as science frames. Our fiducial pipeline reduction does not include the step for resampling spectral data during Stage~3 and instead optimally extracts one-dimensional (1D) spectra for each of the $24$ exposures as \texttt{x1d} files. We run Stage~3 a second time, but include the spectral resampling step since it extracts two-dimensional (2D) spectra for each of the exposures, then combines these to produce the final 2D spectrum as an \texttt{s2d} file (shown in the upper panel of Figure~\ref{fig:Full_Spectrum}) and the final 1D spectrum as a \texttt{x1d} file (shown in the bottom panel of Figure~\ref{fig:Full_Spectrum}); this is the spectrum used for all subsequent analyses.

Figure~\ref{fig:Full_Spectrum} illustrates the final MIRI/LRS 2D and 1D spectra of \mbox{JADES-GS-z14-0}. We provide the 2D spectrum in the upper panel of Figure~\ref{fig:Full_Spectrum} along with the corresponding color bar for the measured signal-to-noise ratio ($\mathrm{S/N}$) per pixel. JWST/MIRI's detector plate scale is $0.11\ \mathrm{arcsec/pixel}$, as shown by the label for the $y$-axis. Blue, \mbox{horizontal} lines indicate the wavelength-dependent profile adopted for the 1D optimal extraction while the blue, vertical lines indicate the observed wavelengths for some of the strongest rest-frame optical emission lines: $[\mathrm{OII}] \lambda\lambda 3727{,}3729$, $\mathrm{H}\beta$, $[\mathrm{OIII}] \lambda\lambda 4959{,}5007$, and $\mathrm{H}\alpha$. The 1D spectrum is provided in the lower panel with the measured flux densities shown by the solid black line and the flux uncertainties ($1\sigma$) shown by the gray shaded region. The quoted uncertainties are derived using Stage~3 of the pipeline and provided by the error extension of the final \texttt{x1d} file, although we implement an additional noise inflation term (see discussion later in this section).

We note that in the first version of this paper, we also indicated the location of $\mathrm{HeI} \lambda 5876$ in Figure~\ref{fig:Full_Spectrum} since this line was surprisingly present in the 1D spectrum, yet appeared in only one nod of our observations. But after acquiring the fourth set of observations, the new data refuted the tentative detection ($\approx 2 \sigma$) of this rarely seen $\mathrm{HeI} \lambda 5876$ line. For this reason, we omit $\mathrm{HeI} \lambda 5876$ from all subsequent discussion in this version of the paper.

There are three spectroscopic features clearly detected above the noise level in both the 2D and 1D spectra of JADES-GS-z14-0 (see Figure~\ref{fig:Full_Spectrum}). These features include the unresolved $[\mathrm{OIII}] \lambda\lambda 4959{,}5007$ doublet detected with $\mathrm{S/N} \approx 14$, unresolved $\mathrm{H}\alpha$ detected with $\mathrm{S/N} \approx 4$, and the rest-frame optical continuum detected with $\mathrm{S/N} > 1$ per wavelength bin at $\lambda_{\mathrm{obs}} \lesssim 6\ \mu\mathrm{m}$ ($\lambda_{\mathrm{rest}} \lesssim 4000\ \mathrm{\AA}$). We note the location of the Balmer continuum limit at $\lambda_{\mathrm{rest}} \approx 3650\ \mathrm{\AA}$ with the gray, vertical line in the lower panel of Figure~\ref{fig:Full_Spectrum}. The stacked continuum is detected at $\mathrm{S/N} \approx 10$ on both sides of the Balmer limit. However, the continuum includes contributions from both \mbox{JADES-GS-z14-0} and a neighboring foreground galaxy. We do not discuss the continuum measurements nor their scientific interpretation in this work as these will be discussed in a forthcoming manuscript (Helton et~al. 2026, \mbox{in preparation}), in which they will be used to infer the stellar population properties of \mbox{JADES-GS-z14-0}.

\begin{figure*}
    \centering
    \includegraphics[width=1.0\linewidth]{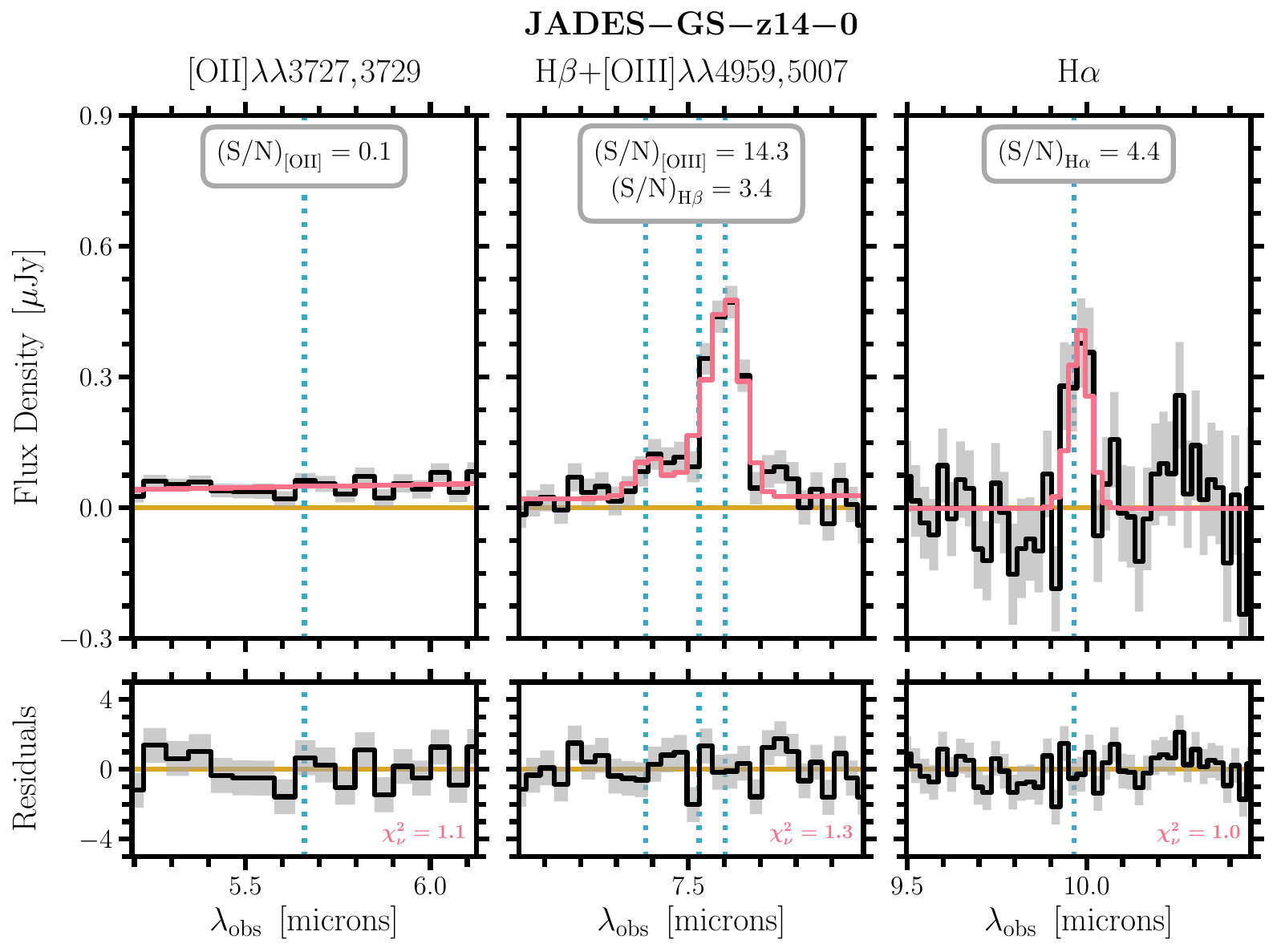}
    \caption{\textbf{Zoomed-in views around the strongest rest-frame optical emission lines.} From left to right, we show the final MIRI/LRS 1D spectrum centered around $[\mathrm{OII}] \lambda\lambda 3727{,}3729$, $\mathrm{H}\beta + [\mathrm{OIII}] \lambda\lambda 4959{,}5007$, and $\mathrm{H}\alpha$, respectively. \textit{Top panels}: The best-fit continua and line profiles are shown in red along with the measured flux densities shown by the black lines and the flux uncertainties ($1\sigma$) shown by the gray shaded regions. Blue, vertical lines once again indicate the locations for some of the strongest emission lines. \textit{Bottom panels}: The residuals of the best-fit models compared with the observations are shown by solid black lines and gray shaded regions. We report the reduced chi-squared statistics to demonstrate the quality of our fits. \label{fig:ZoomIn_EmissionLines}}
\end{figure*}

However, there was some concern that the error extension of the final \texttt{x1d} file was underestimating the true, intrinsic uncertainty of the flux measurements, due to incomplete uncertainty propagation in the aforementioned background subtraction steps. This is a well-known issue\footnote{For more information about known issues with the MIRI/LRS, please also see \href{https://jwst-docs.stsci.edu/known-issues-with-jwst-data/miri-known-issues/miri-lrs-known-issues}{https://jwst-docs.stsci.edu/known-issues-with-jwst-data/miri-known-issues/miri-lrs-known-issues}. We speculate that the underestimated uncertainties for the MIRI/LRS are caused by the pipeline not propagating uncertainty terms related to the global background and influence of cosmic rays. We are also concerned that the flat field for the MIRI/LRS is not well calibrated at such extreme depths.} when using the JWST pipeline to reduce data from the MIRI/LRS. To properly characterize the uncertainties in our combined spectrum, we took advantage of the $24$ exposures that were acquired across the three successful visits and utilized a bootstrap resampling approach to calculate the full covariance matrix. This covariance matrix captures both the variance at each wavelength bin and the correlations between nearby bins. The bootstrap resampling approach provides a robust, empirical estimate of the uncertainty distribution since it makes no assumptions about the underlying noise distribution. For each of the $1000$ bootstrap iterations, we randomly selected exposures with replacement from the full set, computed a mean spectrum using the \texttt{x1d} files from the selected exposures (i.e., the so-called ``subspectra''), and constructed the ensemble of bootstrap realizations. The covariance matrix was finally calculated using the full distribution of these bootstrapped subspectra. The methodology described here has already been used to analyze deep JWST/NIRSpec observations from JADES \citep[see also, e.g.,][]{Hainline:2024, Witstok:2025}. After calculating the full covariance and correlation matrices, we found that the error extension of the final \texttt{x1d} file underestimates uncertainties by roughly $40-45\%$ without any obvious wavelength dependence. To reach this conclusion, we ignored wavelength bins more than five bins removed from one another because the vast majority of correlations between the widely separated bins are consistent with zero. In \mbox{Figures~\ref{fig:Full_Spectrum} and \ref{fig:ZoomIn_EmissionLines}}, we inflate the errors to account for underestimated uncertainties. Our result has broader applicability for other work studying high-redshift galaxies using deep observations with the MIRI/LRS, especially observing programs that have fewer exposures and therefore cannot provide their own empirical estimate of the uncertainties.

We were also concerned about the flux calibration of the MIRI/LRS spectrum due to uncertain slit losses. While \mbox{JADES-GS-z14-0} is smaller than JWST/MIRI's PSF, it is still spatially extended. However, the pipeline performed optimal extraction assuming that it was a point source, since this is the only option for optimal extraction. To check the global flux calibration and test the pipeline's point source assumption, we derived synthetic photometry in the MIRI/F770W filter using the 1D spectrum presented in Figure~\ref{fig:Full_Spectrum} and measured a flux density of $f_{\mathrm{F770W}} \approx 41.5\ \mathrm{nJy}$. We then compared this synthetic photometry with the measured flux density of $f_{\mathrm{F770W}} \approx 74.4 \pm 5.6 \ \mathrm{nJy}$ from \citet[][]{Helton:2025}, taking advantage of the ultra-deep MIRI/F770W imaging from that work, with an on-source integration time of $t_{\mathrm{obs}} \approx 23.8\ \mathrm{hr}$. These two photometric measurements do not agree with one another. However, we additionally compared the average flux density at $\lambda_{\mathrm{obs}} \approx 5.0-5.5\ \mu\mathrm{m}$ of $f \approx 45.8 \pm 5.1\ \mathrm{nJy}$ using the 1D spectrum presented in Figure~\ref{fig:Full_Spectrum} with the measured flux density using the NIRCam/F444W filter of $f_{\mathrm{F444W}} \approx 46.9 \pm 0.6 \ \mathrm{nJy}$ from \citet[][]{Helton:2025}. These two photometric measurements agree with one another. Thus, we reach conflicting conclusions when performing these two different tests, which makes it difficult to determine if slit losses in the MIRI/LRS spectrum presented in this work are substantial. Another complication is that the synthetic photometry includes an uncertain but non-negligible contribution from a nearby foreground galaxy at $z = 3.475$ with NIRCam ID 183349 \citep[][]{Carniani:2024}, while the measured flux density of $f_{\mathrm{F770W}} \approx 74.4 \pm 5.6 \ \mathrm{nJy}$ from \citet[][]{Helton:2025} is for JADES-GS-z14-0 alone (and $f_{\mathrm{F770W}} \approx 46.3 \pm 4.6 \ \mathrm{nJy}$ is measured for NIRCam ID 183349). Compared with a wavelength-independent boxcar extraction, it is more difficult to determine the fraction of light from the neighboring foreground galaxy when using a wavelength-dependent optimal extraction. For a discussion of empirical constraints on contamination in the emission lines, see the end of Section~\ref{SectionFour}.

\section{Analysis of the Emission Lines}
\label{SectionFour}

Figure~\ref{fig:ZoomIn_EmissionLines} provides zoomed-in views around strong rest-frame optical emission lines with the left, \mbox{middle}, and right panels centered around $[\mathrm{OII}] \lambda\lambda 3727{,}3729$, $\mathrm{H}\beta + [\mathrm{OIII}] \lambda\lambda 4959{,}5007$, and $\mathrm{H}\alpha$, respectively. In the upper panels of Figure~\ref{fig:ZoomIn_EmissionLines}, the best-fit continua and line profiles are shown in red along with the measured flux densities shown by the black lines and the flux uncertainties ($1\sigma$) shown by the gray shaded regions. Blue, vertical lines indicate the locations for some of the strongest optical emission lines. These best-fit models do well at reproducing the observations, as demonstrated by the residuals (i.e., $\chi = [ f_{\mathrm{obs}} - f_{\mathrm{model}} ] / \sigma$) shown in the lower panels. The reduced chi-squared statistics are also provided in the lower panels to demonstrate the quality of our fits since these values are all close to one.

We measure fluxes, redshifts, and velocity dispersions for the emission lines shown in Figure~\ref{fig:ZoomIn_EmissionLines} by using \texttt{LMFIT} \citep[][]{Newville:2014} to perform nonlinear least-squares minimization with the Levenberg-Marquardt algorithm. The four spectral windows provided in Figure~\ref{fig:ZoomIn_EmissionLines} are fit separately across the full wavelength range shown in each panel. For each window, we adopt a polynomial for the continuum and a set of Gaussian profiles for the emission lines. For the first of these panels, where the continuum is well detected at $> 1 \sigma$ per wavelength bin, we assume a first-order polynomial. For the other two panels, where the continuum is not well detected, we instead assume a zeroth-order polynomial. Each of the windows from Figure~\ref{fig:ZoomIn_EmissionLines} has a different set of redshifts and velocity dispersions due to uncertain wavelength calibrations and wavelength-dependent resolving powers for the MIRI/LRS \citep[e.g.,][]{Beiler:2023, Xuan:2024}.

Two Gaussians are used to fit the $[\mathrm{OII}] \lambda\lambda 3727{,}3729$ doublet, assuming the flux ratio of the doublet is equal to unity. The precise details of the model do not matter for this fit, since the doublet is completely unresolved; the interline separation ($R \approx 1000$) is more than an order of magnitude smaller than the \mbox{instrumental} line-spread function (LSF) at $\lambda_{\mathrm{obs}} \approx 5.7\ \mu\mathrm{m}$ \citep[e.g.,][]{Kendrew:2016}. Three Gaussians are used to fit $\mathrm{H}\beta$ and the $[\mathrm{OIII}] \lambda\lambda 4959{,}5007$ doublet, assuming the flux ratio of the doublet is equal to $2.98$. A single Gaussian is used to fit $\mathrm{H}\alpha$, assuming negligible contributions from the $[\mathrm{NII}] \lambda\lambda 6548{,}6583$ doublet since these lines are typically at least an order of magnitude weaker than $\mathrm{H}\alpha$ at $z > 3$ \citep[e.g.,][]{Cameron:2023}. A negligible contribution from $[\mathrm{NII}]$ is supported by the nondetection of nitrogen lines in the rest-frame UV along with the nondetection of other nearby low-ionization transitions, such as $\mathrm{[SII]}$. We caution that it is difficult to provide robust emission line measurements for the $[\mathrm{OII}] \lambda\lambda 3727{,}3729$ doublet due to an uncertain local continuum.

\begin{table}
    \centering
    \caption{Measured line fluxes and equivalent widths for the rest-frame optical emission lines of JADES-GS-z14-0. The reported line fluxes have not been corrected for lensing magnification \citep[$\mu = 1.17$;][]{Carniani:2024}. The equivalent widths should be treated as lower limits since the continuum includes an uncertain contribution from a nearby foreground galaxy. We assume the flux ratios of the $[\mathrm{OII}] \lambda\lambda 3727{,}3729$ and $[\mathrm{OIII}] \lambda\lambda 4959{,}5007$ doublets are equal to unity and 2.98, respectively. Despite being lower limits, our spectroscopic measurements of the equivalent widths are larger than those indirectly inferred from the photometry \citep[][]{Helton:2025}.}
    \label{tab:EmissionLineMeasurements}
    \begin{tabular}{c|c|c}
        \hline
        \hline
        & Line Flux & Equivalent Width \\
        Emission Line(s) & [$10^{-19}\,\mathrm{erg/s/cm^{2}}$] & [$\mathrm{\AA}$, Rest-Frame] \\
        \noalign{\vskip 1pt}
        \hline
        $[\mathrm{OII}] \lambda\lambda 3727{,}3729$ & $0.2 \pm 1.8$ & $0 \pm 20$ \\
        $\mathrm{H}\beta$ & $5.3 \pm 1.6$ & $> 140\ (3\sigma)$ \\
        $[\mathrm{OIII}] \lambda 4959$ & $7.7 \pm 0.5$ & $> 220\ (3\sigma)$ \\
        $[\mathrm{OIII}] \lambda 5007$ & $23.0 \pm 1.6$ & $> 650\ (3\sigma)$ \\
        $\mathrm{H}\alpha$ & $8.8 \pm 2.0$ & $> 350\ (3\sigma)$ \\
        \hline
    \end{tabular}
\end{table}

Table~\ref{tab:EmissionLineMeasurements} provides the rest-frame optical emission line fluxes and equivalent widths, as measured in this work using the recently acquired MIRI/LRS spectrum. We measure $z = 14.201 \pm 0.007$ and $\sigma = 1030 \pm 240\ \mathrm{km/s}$ by fitting $\mathrm{H}\beta$ and the $[\mathrm{OIII}] \lambda\lambda 4959{,}5007$ doublet, which are consistent with the previous spectroscopic redshifts (see Table~\ref{tab:PhysicalProperties}) and the expected instrumental resolution at $\lambda_{\mathrm{obs}} \approx 7.5\ \mu\mathrm{m}$ \citep[$R \approx 100$;][]{Kendrew:2016}. The equivalent widths reported in Table~\ref{tab:EmissionLineMeasurements} are quoted as $3\sigma$ lower limits because the continuum is not well detected at $\lambda_{\mathrm{obs}} \gtrsim 6\ \mu\mathrm{m}$. Furthermore, these equivalent widths should be treated as lower limits since the measured continuum includes an uncertain contribution from a neighboring foreground galaxy, as briefly discussed at the end of Section~\ref{SectionThree}. Despite being reported as lower limits, our spectroscopic measurements of the rest-frame \mbox{equivalent} widths ($\mathrm{EW}_{\mathrm{H}\beta \mathrm{+} [\mathrm{OIII}]} > 1010\ \mathrm{\AA},\ 3\sigma$) are still larger than similar values indirectly inferred from the \mbox{photometry} \citep[$\mathrm{EW}_{\mathrm{H}\beta \mathrm{+} [\mathrm{OIII}]} = 370_{-130}^{+360}\ \mathrm{\AA}$;][]{Helton:2025}. These are consistent with but smaller than values for similarly luminous galaxies at $z \approx 8$ \citep[e.g.,][]{Endsley:2024}.

Up to this point, we implicitly assumed that the emission line fluxes are produced by \mbox{JADES-GS-z14-0} alone. However, JADES-GS-z14-0 is close in projection to a foreground galaxy at a separation of roughly $0.4\ \mathrm{arcsec}$ to the east \citep[with NIRCam ID 183349 and a spectroscopic redshift of $z = 3.475$ from emission lines;][]{Carniani:2024}. Deblending the flux from these two galaxies is challenging but essential for the physical interpretation of our observations. Given the size of MIRI/LRS's slit ($4.7\ \mathrm{arcsec}$ in length and $0.51\ \mathrm{arcsec}$ in width; see also Figure~\ref{fig:Slit_Locations}), there will be significant contamination from 183349 in both the emission lines and continuum of \mbox{JADES-GS-z14-0}. Unfortunately, this is not a minor concern since 183349 has rest-frame near-infrared lines that will directly overlap some of \mbox{JADES-GS-z14-0's} rest-frame optical lines. For example, $[\mathrm{FeII}]$ emission lines at rest-frame $1.257\ \mu\mathrm{m}$ and $1.644\ \mu\mathrm{m}$ will directly coincide with \mbox{JADES-GS-z14-0's} $[\mathrm{OII}] \lambda\lambda 3727{,}3729$ and $\mathrm{H}\beta$ lines, respectively. To provide an upper limit on the strengths of 183349's $[\mathrm{FeII}]$ lines, we search for the stronger Paschen lines in the combined MIRI/LRS spectrum of JADES-GS-z14-0 and 183349. Models of star-forming galaxies and AGN suggest that the Paschen lines should always be stronger than the $[\mathrm{FeII}]$ lines \citep[e.g.,][]{Calabro:2023}. For 183349, we expect to see $\mathrm{Pa}\beta$ at $\lambda_{\mathrm{obs}} \approx 5.737\ \mu\mathrm{m}$ and $\mathrm{Pa}\alpha$ at $\lambda_{\mathrm{obs}} \approx 8.391\ \mu\mathrm{m}$. A gray, vertical line indicates the location of $\mathrm{Pa}\alpha$ in the lower panel of \mbox{Figure~\ref{fig:Full_Spectrum}}. Since there is negative flux in the 1D spectrum at this location, we find no evidence for $\mathrm{Pa}\alpha$; this indicates that the measured emission line fluxes are from JADES-GS-z14-0 alone, with a negligible contribution from the neighboring foreground galaxy.

\begin{table*}
    \caption{Empirical and inferred physical properties of JADES-GS-z14-0.}
    \label{tab:PhysicalProperties}
    \begin{threeparttable}
        \renewcommand{\arraystretch}{1.05}
        \makebox[\textwidth]{
        \hspace*{-10mm}
        \begin{tabular}{lr}
            \hline
            \hline
            \multicolumn{2}{c}{Empirical Properties} \\
            \hline
		      R.A. [degrees (ICRS)] & $+53.08294$ \\
		      Decl. [degrees (ICRS)] & $-27.85563$ \\
            Spectroscopic Redshift ($z_{\mathrm{Ly}\alpha}$)\tnote{a} & $14.32_{-0.20}^{+0.08}$ \\
            Spectroscopic Redshift ($z_{\mathrm{CIII]} \lambda\lambda 1907{,}1909}$)\tnote{a} & $14.178 \pm 0.013$ \\
            Spectroscopic Redshift ($z_{\mathrm{[OIII]} \lambda88 \,\mu\mathrm{m}}$)\tnote{b} & $14.1796 \pm 0.0007$ \\
            UV Luminosity ($M_{\mathrm{UV}}$)\tnote{\textdagger,\,a} & $-20.81 \pm 0.16$ \\
            UV Continuum Slope ($\beta_{\mathrm{UV}}$)\tnote{a} & $-2.20 \pm 0.07$ \\
            UV Half-Light Radius ($r_{\mathrm{UV}}/\mathrm{pc}$)\tnote{a} & $260 \pm 20$ \\
            UV Half-Light Radius ($r_{\mathrm{UV}}/\mathrm{arcsec}$)\tnote{a} & $0.079 \pm 0.006$ \\
            Apparent Magnitude ($m_{\mathrm{F277W}} / \mathrm{mag}$)\tnote{c} & $27.05 \pm 0.01$ \\
            Apparent Magnitude ($m_{\mathrm{F444W}} / \mathrm{mag}$)\tnote{c} & $27.22 \pm 0.01$ \\
            Apparent Magnitude ($m_{\mathrm{F770W}} / \mathrm{mag}$)\tnote{c} & $26.72 \pm 0.08$ \\
            \noalign{\vskip 1pt}
            \hline
            \hline
            \multicolumn{2}{c}{Inferred Properties from Standard Strong-Line Diagnostics (Section~\ref{SectionFive})} \\
            \hline
            Diffuse Dust Attenuation ($A_{V}/\mathrm{mag}$)\tnote{d} & $-1.9 \pm 1.3$ \\
            Star Formation Rate ($\mathrm{SFR}/[M_{\odot}/\mathrm{yr}]$)\tnote{\textdagger,\,d} & $8.4 \pm 2.0$ \\
            SFR Surface Density ($\Sigma_{\mathrm{SFR}}/[M_{\odot}/\mathrm{yr}/\mathrm{kpc}^{2}]$)\tnote{\textdagger,\,d} & $19.8 \pm 4.8$ \\
            Ionizing Photon Production Efficiency ($\mathrm{log}_{10} [\xi_{\mathrm{ion}} / \{\mathrm{Hz/erg}\}]$)\tnote{d} & $25.25 \pm 0.12$ \\
            Gas-Phase Oxygen Abundance ($[\mathrm{O/H}]/\mathrm{dex}$)\tnote{d} & $-1.20 \pm 0.23$ \\
            Carbon-to-Oxygen Ratio ($[\mathrm{C/O}]/\mathrm{dex}$)\tnote{d} & $-0.37 \pm 0.24$ \\
            Ionization Parameter ($\mathrm{log}_{10} [U]$)\tnote{d} & $> -2.35\ (3\sigma\ \mathrm{limit})$ \\
            Electron Density ($n_{e}/\mathrm{cm}^{-3}$)\tnote{d} & $690 \pm 200$ \\
            \noalign{\vskip 1pt}
            \hline
            \hline
            \multicolumn{2}{c}{Inferred Properties from Detailed Photoionization Modeling (Section~\ref{SectionSix})} \\
            \hline
            Ionizing Photon Production Efficiency ($\mathrm{log}_{10} [\xi_{\mathrm{ion}} / \{\mathrm{Hz/erg}\}]$)\tnote{d} & $25.25_{-0.09}^{+0.08}$ \\
            Gas-Phase Oxygen Abundance ($[\mathrm{O/H}]/\mathrm{dex}$)\tnote{d} & $-0.24_{-0.41}^{+0.38}$ \\
            Nitrogen-to-Oxygen Ratio ($[\mathrm{N/O}]/\mathrm{dex}$)\tnote{d} & $-0.28_{-0.49}^{+0.57}$ \\
            Carbon-to-Oxygen Ratio ($[\mathrm{C/O}]/\mathrm{dex}$)\tnote{d} & $+0.08_{-0.41}^{+0.42}$ \\
            Ionization Parameter ($\mathrm{log}_{10} [U]$)\tnote{d} & $-1.35_{-0.36}^{+0.25}$ \\
            Electron Density ($n_{e}/\mathrm{cm}^{-3}$)\tnote{d} & $540_{-320}^{+520}$ \\
            \noalign{\vskip 1pt}
            \hline
            \hline
            \multicolumn{2}{c}{Inferred Properties from SED Fitting \citep[][]{Helton:2025}} \\
            \hline
            Stellar Mass ($\mathrm{log}_{10}[M_{\ast}/M_{\odot}]$)\tnote{\textdagger,\,c} & $8.72_{-0.40}^{+0.44}$ \\
            Stellar Metallicity ($\mathrm{log}_{10}[Z_{\ast}/Z_{\odot}]$)\tnote{c} & $-1.84_{-0.81}^{+0.66}$ \\
            Mass-Weighted Stellar Age ($t_{\ast}/\mathrm{Myr}$)\tnote{c} & $15.2_{-12.5}^{+28.1}$ \\
            Star Formation Rate ($\mathrm{SFR}_{10} / [M_{\odot}/\mathrm{yr}]$)\tnote{\textdagger,\,c} & $25.1_{-5.6}^{+5.4}$ \\
            SFR Surface Density ($\Sigma_{\mathrm{SFR}10} / [M_{\odot}/\mathrm{yr}/\mathrm{pc}^{2}]$)\tnote{\textdagger,\,c} & $64_{-14}^{+14}$ \\
            Optical Equivalent Width ($\mathrm{EW}_{\mathrm{[OIII]} + \mathrm{H}\beta} / \mathrm{\AA}$)\tnote{c} & $370_{-130}^{+360}$ \\
            Diffuse Dust Attenuation ($A_{V} / \mathrm{mag}$)\tnote{c} & $0.557_{-0.131}^{+0.081}$ \\
            \noalign{\vskip 1pt}
            \hline
            \hline
            \multicolumn{2}{c}{Other Inferred Properties} \\
            \hline
            Neutral Hydrogen Column Density ($\mathrm{log}_{10}[N_{\mathrm{HI}} / \mathrm{cm}^{-2}]$)\tnote{a} & $22.23_{-0.08}^{+0.08}$ \\
            Neutral Hydrogen Column Density ($\mathrm{log}_{10}[N_{\mathrm{HI}} / \mathrm{cm}^{-2}]$)\tnote{b} & $21.96_{-0.09}^{+0.08}$ \\
            Dynamical Mass ($\mathrm{log}_{10}[M_{\mathrm{dyn}}/M_{\odot}]$)\tnote{\textdagger,\,b} & $9.0_{-0.2}^{+0.2}$ \\
            Dynamical Mass ($\mathrm{log}_{10}[M_{\mathrm{dyn}}/M_{\odot}]$)\tnote{\textdagger,\,e} & $9.4_{-0.4}^{+0.8}$ \\
            Rotational Support ($V_{\mathrm{rot}} / \sigma$)\tnote{e} & $> 2.5$ \\
            \hline
	\end{tabular}
    }
	\begin{tablenotes}
	    \footnotesize
            \item \textbf{Notes.}
            \item[\textdagger] \mbox{This quantity has been corrected for a lensing magnification of $\mu = 1.17$ \citep[][]{Carniani:2024}.}
            \item[a] \citet[][]{Carniani:2024}.
            \item[b] \citet[][]{Carniani:2025}.
            \item[c] \citet[][]{Helton:2025}.
            \item[d] This work.
            \item[e] \citet[][]{Scholtz:2025a}.
    \end{tablenotes}
    \end{threeparttable}
\end{table*}

\section{Inferring Physical Properties Using Standard Strong-Line Diagnostics}
\label{SectionFive}

As already mentioned, the rest-frame optical emission lines are highly diagnostic since they encode valuable \mbox{information} about the incident ionizing spectrum, which also includes the detailed properties of the stellar populations and interstellar medium. In this section, we explore the physical properties of \mbox{JADES-GS-z14-0} as inferred from strong-line diagnostics that primarily use the brightest rest-frame optical emission lines. These include the diffuse dust attenuation (Section~\ref{SectionFiveOne}), recent star formation rate (Section~\ref{SectionFiveTwo}), ionizing photon production efficiency (Section~\ref{SectionFiveThree}), gas-phase oxygen abundance (Section~\ref{SectionFiveFour}), ionization parameter (Section~\ref{SectionFiveFive}), and electron density (Section~\ref{SectionFiveSix}). The empirical and inferred properties of \mbox{JADES-GS-z14-0} are succinctly summarized in Table~\ref{tab:PhysicalProperties}, with many of the values coming from this work's emission line modeling.

To place the physical properties of \mbox{JADES-GS-z14-0} in the context of other star-forming galaxies at high redshifts, we compare with galaxies from the JADES Data Release 4 \citep[DR4;][]{Curtis-Lake:2026, Scholtz:2026}, which includes the complete JWST/NIRSpec spectroscopy of JADES. For this work, we use measurements from the five-pixel extraction of the low-resolution NIRSpec/PRISM spectrum. We select $N \approx 1500$ galaxies at $z > 3$ with spectroscopic redshifts from at least one emission line detection ($\texttt{z\_Spec\_Flag} ==\,\texttt{A},\ \texttt{B},\ \mathrm{or}\ \texttt{C}$). For galaxies with multiple entries in the JADES DR4, we select the entry with the longest exposure time. These galaxies are represented by gray circles throughout the manuscript. We further select $N \approx 850$ galaxies at $z > 3$ with high-quality redshifts from the previous comparison sample by requiring $\mathrm{H}\beta$ to have $\mathrm{S/N} > 4$. These galaxies are represented by green squares throughout. By requiring $\mathrm{H}\beta$ to be well detected, we are effectively making a cut on the recent star formation rate.

We additionally compare with three galaxies at $z > 10$ from the literature that have spectroscopic redshifts from at least one emission line detection and previous MIRI/Spectroscopy. In order of decreasing redshift, these three galaxies include \mbox{GHZ2/GLASSz12} at $z = 12.34$ \citep[]{Castellano:2024, Zavala:2025}, \mbox{GN-z11} at $z = 10.60$ \citep[][]{Bunker:2023, Alvarez-Marquez:2025}, and \mbox{MACS0647-JD1} at $z = 10.17$ \citep[][]{Hsiao:2024a, Hsiao:2024b}. Similar to the observations of JADES-GS-z14-0 presented in this work, the previous observations of MIRI/Spectroscopy have targeted some of the brightest rest-frame optical nebular emission lines. For self-consistency, we rederive physical properties for each of the galaxies in the comparison samples using the same methodology as used for JADES-GS-z14-0.

\subsection{Diffuse Dust Attenuation}
\label{SectionFiveOne}

Existing predictions for the diffuse dust attenuation \citep[$A_{V} \approx 0.2-0.6\ \mathrm{mag}$;][]{Carniani:2024, Carniani:2025, Helton:2025} suggest that the measured line fluxes should be unaffected by dust, at least within the quoted uncertainties. We can directly check this with the Balmer decrement (i.e., $\mathrm{H} \alpha / \mathrm{H} \beta$) since any deviation from its intrinsic value can be attributed to dust attenuation. An intrinsic value of $2.86$ is appropriate for gas with an electron density of $n_{\mathrm{H}} = 10^{2}\ \mathrm{cm}^{-3}$ and an electron temperature of $T_{e} = 10^{4}\ \mathrm{K}$ when assuming Case~B recombination. The assumed density is consistent with the upper limit from the inferred $\mathrm{[OIII]} \lambda 5007 / \mathrm{[OIII]} \lambda88 \mu\mathrm{m}$ ratio ($n_{\mathrm{H}} < 700\ \mathrm{cm}^{-3}$), which was estimated using single-zone photoionization models \citep[][]{Carniani:2025}, but see also the discussion of densities in Section~\ref{SectionFiveSix} and Section~\ref{SectionSix}. However, we note that the Balmer decrement is largely insensitive to density and only marginally sensitive to temperature. We measure $\mathrm{H} \alpha / \mathrm{H} \beta = 1.7 \pm 0.6$, which is smaller than the intrinsic value predicted by Case~B recombination and thus consistent with no measurable dust attenuation in \mbox{JADES-GS-z14-0}, so we assume $A_{V} = 0$ for the remainder of this manuscript. Our simplifying assumption is the same one used to study the other galaxies at $z > 10$ with previous spectroscopy from JWST/MIRI \citep[e.g.,][]{Zavala:2025}. However, we note that if we assume the attenuation curve from \citet[][]{Calzetti:2000} and then use our measured Balmer decrement to infer the $V$-band dust attenuation, we measure a value of $A_{V} \approx -1.9 \pm 1.3\ \mathrm{mag}$.

Although $\mathrm{H}\beta$ is only marginally detected ($\approx 3 \sigma$), our measurement of the Balmer decrement indicates a relatively low dust content in the interstellar medium of \mbox{JADES-GS-z14-0} when compared to similarly massive galaxies ($M_{\ast} \approx 10^{8}-10^{9}\ M_{\odot}$) at lower redshifts ($z \approx 6$). \citet{Carniani:2024} discuss the low dust content of \mbox{JADES-GS-z14-0} in great detail, and they find that the observed dust attenuation can possibly be explained by three different scenarios: ($\#1$) large amounts of dust are distributed on large scales due to galactic outflows, ($\#2$) dust compositions are different from at lower redshifts, and/or ($\#3$) the destruction rate of dust grains from supernova shock waves is higher than expected.

\citet[][]{Schneider:2024} reviewed the formation and evolution of dust in the early Universe while focusing on the different sources of dust grains. \mbox{They argue} that dust grains produced by core-collapse supernovae represent the most likely formation pathway for dust in the first billion years of cosmic time \citep[$z > 5$; see also, e.g.,][]{McKinney:2025}. Their argument revolves around the fact that other formation channels for dust grains that are dominant in lower-redshift galaxies, including asymptotic giant branch (AGB) stars and ambient grain growth in the cold interstellar medium, have timescales that are comparable to or greater than the age of the Universe for galaxies at $z > 5$. However, ambient grain growth may still be needed to explain the large observed dust masses in galaxies at $z \approx 6$ \citep[][]{Narayanan:2026}. Core-collapse supernovae produce fewer small grains than other formation channels, since only larger grains are capable of surviving the reverse shock of supernovae in order to reach the interstellar medium. This implies that attenuation curves should flatten at high redshifts with only a weak dependence on wavelength, which has important implications for the inferred properties of distant galaxies. For example, with flattened attenuation curves, galaxies can have considerable dust attenuation ($A_{V} \approx 1\ \mathrm{mag}$) without leading to unphysical rest-frame UV colors and stellar masses. Thus, flat attenuation curves would reduce our ability to constrain the dust content of galaxies in the early Universe.

\subsection{Recent Star Formation Rate}
\label{SectionFiveTwo}

\begin{figure*}[t!]
    \centering
    \includegraphics[width=0.9\linewidth]{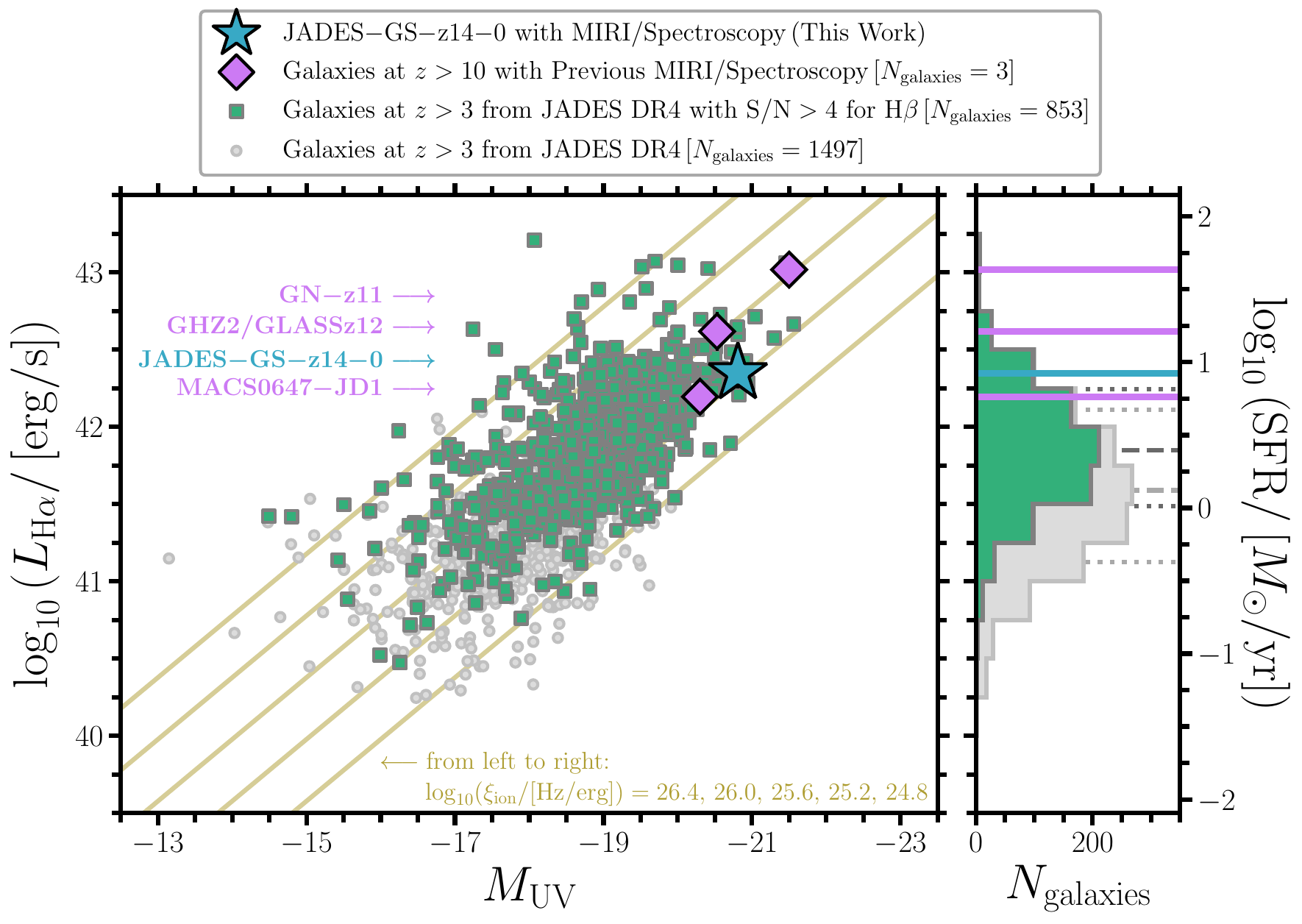}
    \caption{\textbf{\boldmath$\mathrm{H}\alpha$ line luminosity versus absolute UV magnitude.} \textit{Left panel:} These measurements are shown for a few different samples of high-redshift star-forming galaxies at $z > 3$. These include $N \approx 1500$ galaxies at $z > 3$ from the JADES DR4, represented by the gray circles, and $N \approx 850$ of those same galaxies with $\mathrm{H}\beta$ well-detected at $> 4 \sigma$, represented by the green squares \citep[][]{Curtis-Lake:2026, Scholtz:2026}. For galaxies at $z \gtrsim 6$ in the JADES DR4, $\mathrm{H}\alpha$ has shifted beyond JWST/NIRSpec's wavelength coverage, so we use the $\mathrm{H}\beta$ line luminosities instead and assume zero dust attenuation, consistent with Case~B recombination. Three galaxies at $z > 10$ from the literature with previous MIRI/Spectroscopy are illustrated by the purple diamonds. These include, in order of decreasing brightness, GN-z11 at $z = 10.60$ \citep[][]{Bunker:2023, Alvarez-Marquez:2025}, GHZ2/GLASSz12 at $z = 12.34$ \citep[]{Castellano:2024, Zavala:2025}, and MACS0647-JD1 at $z = 10.17$ \citep[][]{Hsiao:2024a, Hsiao:2024b}. Our measurements for JADES-GS-z14-0 at $z = 14.18$ are depicted by the blue star and represent the most distant detection of $\mathrm{H}\alpha$. As a point of comparison, we provide brown lines to show the relation between $L_{\mathrm{H}\alpha}$ and $M_{\mathrm{UV}}$ as parametrized by the ionizing photon production efficiency. \textit{Right panel:} Histograms showing the distributions of $\mathrm{H}\alpha$ line luminosities, which are used as a proxy for the recent star formation rate. Solid lines represent measurements for galaxies at $z > 10$ while gray dashed and dotted lines represent medians and $68\%$ confidence intervals for the samples of galaxies at $z > 3$ from JADES DR4. JADES-GS-z14-0 is among the most actively star-forming galaxies at $z > 3$ with spectroscopic confirmation. \label{fig:LHalpha_vs_MUV}}
\end{figure*}

A galaxy's star formation rate reflects the complex balance between gas supply (through the interplay of gas inflows and outflows) and the efficiency of this gas turning into stars (through the star formation efficiency, or SFE). Observationally, star formation rates can be inferred using multiple semi-independent tracers, each with distinct advantages and limitations. The rest-frame UV continuum is sensitive to the emission from hot, massive, short-lived stars (i.e., O and B stars) and thus probes star formation on timescales of $t \lesssim 100\ \mathrm{Myr}$. On the other hand, the rest-frame optical emission lines, such as the Balmer hydrogen line $\mathrm{H}\alpha$, are sensitive to ionizing radiation from the youngest stellar populations (O stars) and thus probe star formation on shorter timescales of $t \lesssim 10\ \mathrm{Myr}$. Importantly, optical emission is less affected by dust attenuation than UV emission. Additionally, comparing star formation rates that probe different timescales, such as those inferred from the UV continuum emission and the optical emission lines, can provide an important constraint on the burstiness of the recent star formation history (SFH).

\citet[][]{Kennicutt:1998} proposed the classical calibration to convert $\mathrm{H}\alpha$ line luminosities into star formation rates, as demonstrated by Equation~\ref{eq:Ha_to_SFR_conversion_v1}, assuming solar metallicity and a constant SFH over the previous $100\ \mathrm{Myr}$ based on the typical conditions of galaxies in the local Universe \citep[for a review, see][]{Kennicutt:2012}. However, distant star-forming galaxies are known to experience lower metallicities and burstier SFHs than their local counterparts, indicating that the classical calibration should evolve with redshift \citep[e.g.,][]{Shapley:2023}. \citet[][]{Theios:2019} derive updated calibrations for this relation at high redshifts, finding $C_{\mathrm{H}\alpha} = -41.35$ for a Kroupa-type initial mass function \citep[IMF;][]{Kroupa:2001} and solar metallicity, along with a calibration of $C_{\mathrm{H}\alpha} = -41.64$ for sub-solar metallicity. These updated calibrations include binarity in the stellar populations and implicitly assume an escape fraction of zero. \begin{equation}
    \label{eq:Ha_to_SFR_conversion_v1}
    \mathrm{log}_{10} \left( \frac{\mathrm{SFR}}{M_{\odot}/\mathrm{yr}} \right) = \mathrm{log}_{10} \left( \frac{L_{\mathrm{H}\alpha}}{\mathrm{erg/s}} \right) + C_{\mathrm{H}\alpha}
\end{equation}

Finally, \citet[][]{Kramarenko:2026} derive another set of updated calibrations for this relation, as demonstrated by Equation~\ref{eq:Ha_to_SFR_conversion_v2}, finding values between the two quoted by \citet[][]{Theios:2019}. \citet[][]{Kramarenko:2026} include a secondary dependence on the $\mathrm{H}\alpha$ line luminosities since low-mass, low-metallicity galaxies are more efficient at producing $\mathrm{H}\alpha$ photons per unit star formation rate when compared to their higher-mass, higher-metallicity counterparts. They derive their updated calibrations using the SPHINX cosmological simulations to select a sample of star-forming galaxies that are representative of the $\mathrm{H}\alpha$-emitting population observed with JWST. However, at the highest redshifts (i.e., $z > 10$), it is unclear if these calibrations are applicable since the ionizing photon outputs of low-metallicity stars with early Universe abundance patterns are poorly understood. We should also note that the choice of IMF slope and stellar population synthesis model results in a cumulative systematic error of $\approx 0.2-0.3\ \mathrm{dex}$ \citep[][]{Kramarenko:2026}. Varying the high-mass cutoff for the IMF results in an even larger systematic error. Despite these caveats, our work represents one of the first attempts at exploring the applicability of classical calibrations for measuring star formation rates at the redshift frontier. \begin{equation}
    \label{eq:Ha_to_SFR_conversion_v2}
    \begin{aligned}
        \mathrm{log}_{10} \left( \frac{\mathrm{SFR}}{M_{\odot}/\mathrm{yr}} \right) = \mathrm{log}_{10} \left( \frac{L_{\mathrm{H}\alpha}}{\mathrm{erg/s}} \right) -41.45 \\ + 0.06 \left[ \mathrm{log}_{10} \left( \frac{L_{\mathrm{H}\alpha}}{\mathrm{erg/s}} \right) - 41.90 \right]
    \end{aligned}
\end{equation}

For JADES-GS-z14-0, we measure an $\mathrm{H}\alpha$ luminosity of $L_{\mathrm{H}\alpha} = 2.23 \pm 0.50 \times 10^{42}\ \mathrm{erg/s}$ from the measured $\mathrm{H}\alpha$ line flux after accounting for the luminosity distance at $z = 14.18$ and the lensing magnification of $\mu = 1.17$ \citep[][]{Carniani:2024}. We then use this measurement to infer an $\mathrm{SFR} = 8.4 \pm 2.0\ M_{\odot}/\mathrm{yr}$ from Equation~\ref{eq:Ha_to_SFR_conversion_v2}. We further estimate $\Sigma_{\mathrm{SFR}} = 19.8 \pm 4.8\ M_{\odot}/\mathrm{yr}/\mathrm{kpc}^{2}$ using the measured half-light radius ($r_{\mathrm{UV}}$; see also Table~\ref{tab:PhysicalProperties}) from \citet[][]{Carniani:2024} and Equation~\ref{eq:SFR_surface_density}. \begin{equation}
    \label{eq:SFR_surface_density}
    \Sigma_{\mathrm{SFR}} = \frac{\mathrm{SFR}}{2 \pi \left( r_{\mathrm{UV}} \right)^{2}}
\end{equation}

In Figure~\ref{fig:LHalpha_vs_MUV}, we compare $\mathrm{H}\alpha$ luminosities with rest-frame UV magnitudes for various samples of star-forming galaxies at $z > 3$. \mbox{JADES-GS-z14-0} is among the most actively star-forming galaxies at $z > 3$ with spectroscopic confirmation.

Despite being spatially extended and among the largest galaxies observed at $z > 10$, our measurements confirm that JADES-GS-z14-0 has a star formation rate surface density that is comparable to some of the most vigorous starbursts observed in the local Universe \citep[e.g.,][]{Kennicutt:2021}. This is an important result to consider because, at such extreme surface densities, it is predicted that the escape fraction of ionizing photons should be nonzero and possibly approaching order unity \citep[e.g.,][]{Sharma:2016, Izotov:2018}. A nonzero escape fraction would imply that the star formation rates derived from emission lines in this work are underestimates of the true values. The escape fraction is discussed in a bit more detail in Section~\ref{SectionFiveThree}. We will also discuss the escape fraction as inferred from fitting the panchromatic spectral energy distribution (SED) in a forthcoming manuscript from the JADES collaboration (Helton et~al. 2026, \mbox{in preparation}).

Our measurements of the recent star formation rate and star formation rate surface density are significantly smaller than similar values inferred from SED fitting that did not include the newly acquired MIRI/LRS data \citep[][]{Carniani:2024, Carniani:2025, Helton:2025}. These previous works estimated \mbox{$\mathrm{SFR}_{10} \approx 15-30\ M_{\odot}/\mathrm{yr}$} by averaging the inferred SFH over the most recent $10\ \mathrm{Myr}$ of lookback time. We speculate that this discrepancy is caused by \mbox{JADES-GS-z14-0} experiencing a lull (or decline) in the most recent few million years of the SFH, thereby affecting the shorter timescale emission lines more than the longer timescale UV continuum emission. This explanation is consistent with the SFH inferred by \citet[][]{Carniani:2025}. We return to this idea of a lulling (or declining) SFH for JADES-GS-z14-0 in Section~\ref{SectionSeven}.

\subsection{Ionizing Photon Production Efficiency}
\label{SectionFiveThree}

\begin{figure*}[t!]
    \centering
    \includegraphics[width=0.9\linewidth]{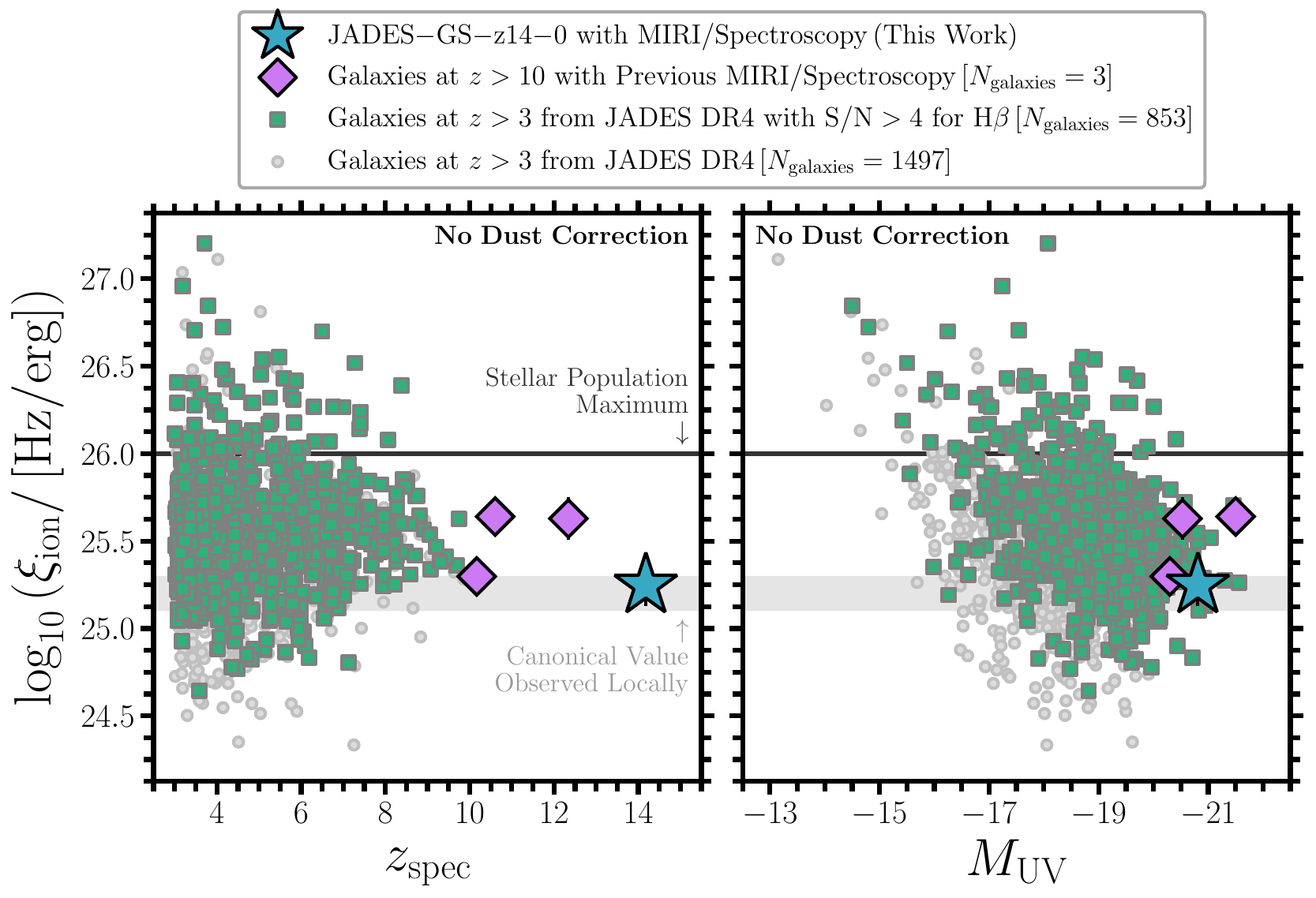}
    \caption{\textbf{Ionizing photon production efficiency versus both spectroscopic redshift and absolute UV magnitude.} \textit{Left panel:} We show the same three samples of star-forming galaxies at $z > 3$ from Figure~\ref{fig:LHalpha_vs_MUV} using a consistent plotting scheme. For comparison, the canonical values that have been predicted and observed in galaxies at $z \lesssim 4$ are shown by the gray shaded region \citep[e.g.,][]{Madau:1999, Bouwens:2016} while the approximate stellar population maximum is shown by the black line \citep[e.g.,][]{Schaerer:2025}. Among the four galaxies at $z > 10$ with these measurements, there are three that are similarly efficient at producing ionizing photons and include JADES-GS-z14-0, GHZ2/GLASSz12, and GN-z11. No dust corrections have been applied to these galaxies. The observed redshift evolution in the lower envelope of green points is caused by observational limitations and burstier SFHs at high redshifts, while the evolution in the upper envelope is caused by smaller dust reservoirs at high redshifts. \textit{Right panel:} The faintest galaxies are observed as the most efficient producers of ionizing photons.  \label{fig:xi_ion_vs_zSpec_and_MUV}}
\end{figure*}

Immediately after the formation of the first stars and galaxies, Lyman continuum photons with $\lambda_{\mathrm{rest}} \lesssim 912\ \mathrm{\AA}$ started ionizing the neutral hydrogen that permeated the intergalactic medium (IGM). Equation~\ref{eq:number_of_ionizing_photons} shows that the number of ionizing photons available to reionize the neutral IGM per unit time and comoving volume ($\dot{n}_{\mathrm{ion}}$) is dependent on the abundance of galaxies per unit comoving volume ($\rho_{\mathrm{UV}}$), the \mbox{ionizing} photon production efficiency ($\xi_{\mathrm{ion}}$), and the ionizing photon escape fraction ($f_{\mathrm{esc}}$). Assuming \mbox{canonical} values for the ionizing photon production efficiency, relatively large average escape fractions ($f_{\mathrm{esc}} \approx 10-20\%$) are necessary for galaxies to be the dominant source of reionization \citep[e.g.,][]{Ouchi:2009, Robertson:2013, Robertson:2015, Finkelstein:2019, Naidu:2020}. Both standard models of reionization and observations of galaxies at $z \lesssim 4$ suggest canonical values of $\xi_{\mathrm{ion}} \approx 10^{25.1-25.3}\ \mathrm{Hz/erg}$ \citep[e.g.,][]{Madau:1999, Bouwens:2016}. \begin{equation}
    \label{eq:number_of_ionizing_photons}
    \dot{n}_{\mathrm{ion}} = \rho_{\mathrm{UV}} \xi_{\mathrm{ion}} f_{\mathrm{esc}}
\end{equation}

As described in Equation~\ref{eq:ionizing_photon_production_efficiency}, it is possible to \mbox{indirectly} infer the number of ionizing photons from the \mbox{hydrogen} recombination lines, since those lines are produced \mbox{after} photoionization has occurred. We should note that this is modulo the fraction of ionizing photons absorbed by dust, which can be non-negligible for high ionization parameters \citep[e.g.,][]{Mathis:1986}. $L_{\mathrm{H}\alpha}$ represents the dust-corrected  luminosity of $\mathrm{H}\alpha$ in units of $\mathrm{erg/s}$ while $L_{\mathrm{UV}}$ represents the dust-corrected monochromatic luminosity of the UV in units of $\mathrm{erg/s/Hz}$, as measured at rest-frame $1500\ \mathrm{\AA}$. The adopted calibration is provided by \citet[][]{Osterbrock:2006}, assuming a temperature of $T_{e} = 10^{4}\ \mathrm{K}$ and a density of $n_{\mathrm{H}} = 100\ \mathrm{cm}^{-3}$. This calibration does not account for further complications such as the burstiness of the SFH, IMF, metallicity, and the way in which individual stars evolve (e.g., binarity). \begin{equation}
    \label{eq:ionizing_photon_production_efficiency}
    \xi_{\mathrm{ion}} = \frac{\dot{n}_{\mathrm{ion}}}{L_{\mathrm{UV}}} = \frac{L_{\mathrm{H}\alpha} \left( 7.28 \times 10^{11} \right)}{L_{\mathrm{UV}} \left( 1 - f_{\mathrm{esc}} \right)}
\end{equation}

It is well established that the ionizing photon production efficiency increases with redshift \citep[e.g.,][]{Bouwens:2016, Matthee:2017}. The observed redshift evolution in the ionizing photon production efficiency is likely physical in origin (rather than an \mbox{observational} bias) and possibly caused by the increasing prevalence of low-mass galaxies with bursty SFHs at high redshifts \citep[e.g.,][]{Simmonds:2024a, Simmonds:2024b}. This result is illustrated by Figure~\ref{fig:xi_ion_vs_zSpec_and_MUV}, where we compare ionizing photon production efficiencies with both spectroscopic redshifts and absolute UV magnitudes. The lower envelope of green and gray points, representing galaxies at $z > 3$ from JADES DR4, demonstrates the redshift evolution; the upper envelope demonstrates the increasing importance of dust at lower redshifts. For \mbox{JADES-GS-z14-0}, we infer $\xi_{\mathrm{ion}} = 10^{25.25 \pm 0.12}\ \mathrm{Hz/erg}$ by assuming an escape fraction of zero and zero dust attenuation, since we measure a Balmer decrement consistent with Case~B recombination (see Section~\ref{SectionFiveOne}). Any dust effects would decrease our estimates of the ionizing photon production efficiency, since the rest-frame UV emission should be more attenuated by dust than the rest-frame optical emission, assuming standard attenuation laws.

On the other hand, nonzero escape fractions would increase our estimates of the ionizing photon production efficiency (see Equation~\ref{eq:ionizing_photon_production_efficiency}). \citet[][]{Carniani:2025} indirectly infer $f_{\mathrm{esc}} \approx 4-20\%$ for JADES-GS-z14-0 from SED fitting of the available JWST and ALMA observations, but not including the newly acquired MIRI/LRS data. A nonzero escape fraction was needed to explain the relative weakness of the $\mathrm{CIII]} \lambda\lambda 1907{,}1909$ equivalent width when compared to the strength of $[\mathrm{OIII}] \lambda 88 \mu\mathrm{m}$. An alternative estimate of the escape fraction relies on converting the measured rest-frame UV continuum slope ($\beta_{\mathrm{UV}}$) into an escape fraction by adopting the relation from \citet[][]{Chisholm:2022}. Using their relation, we indirectly infer $f_{\mathrm{esc}} \approx 2-20\%$ for \mbox{JADES-GS-z14-0}. We should caution the reader that the relation from \citet[][]{Chisholm:2022} was calibrated with local ($z \approx 0$) Lyman continuum leaking galaxies and thus their applicability at $z > 10$ is not well understood. \begin{equation}
    \label{eq:escape_fraction}
    f_{\mathrm{esc}} = \left( 1.3 \pm 0.6 \right) \times 10^{- 4 - \beta_{\mathrm{UV}} \left( 1.2 \pm 0.1 \right)}
\end{equation}

If real, the origin of a nonzero escape fraction could be strong galactic outflows that expel gas and dust from the interstellar medium of \mbox{JADES-GS-z14-0}, which has been suggested by \citet[][]{Ferrara:2025} in their ``attenuation-free models'' to possibly explain the surprising abundance of luminous galaxies observed at high redshifts. Their argument is that supernova-driven winds could evacuate a pathway out of the galaxy, allowing ionizing photons to escape. The damped Lyman-$\alpha$ absorption inferred for \mbox{JADES-GS-z14-0} \citep[see Table~\ref{tab:PhysicalProperties};][]{Carniani:2024, Carniani:2025}, which is required to explain the redshifted and smoothed Lyman-$\alpha$ break relative to the emission lines, could be related to a potential galactic outflow.

\subsection{Gas-Phase Oxygen Abundance}
\label{SectionFiveFour}

\begin{figure*}[t!]
    \centering
    \includegraphics[width=0.9\linewidth]{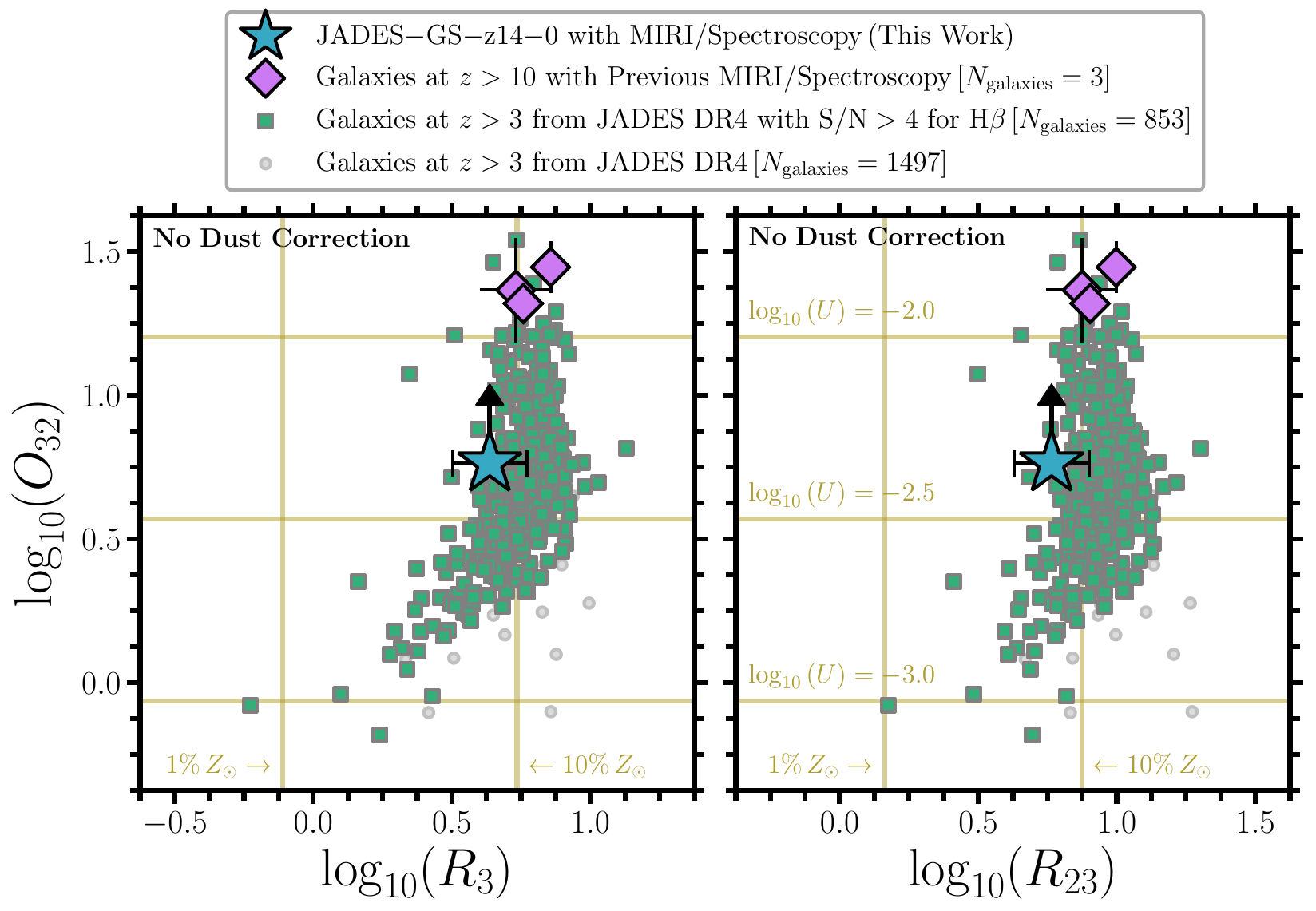}
    \caption{\textbf{Standard strong-line diagnostics with \boldmath$O_{32}$ versus \boldmath$R_{3}$ and \boldmath$R_{23}$.} \textit{Both panels:} These emission line diagnostics are commonly used to infer gas-phase metallicity with $R_{3}$ and $R_{23}$ and ionization parameter with $O_{32}$. We show the same three samples of star-forming galaxies at $z > 3$ from Figure~\ref{fig:LHalpha_vs_MUV}. We also show brown vertical lines of constant metallicity and brown horizontal lines of constant ionization parameter. No dust corrections have been applied to any of these galaxies. We find that JADES-GS-z14-0 and the other three galaxies at $z > 10$ are characterized by extreme ionization conditions and relatively high metallicities, indicating rapid metal enrichment. Simulations struggle to reproduce these results (see the discussion in Section~\ref{SectionSeven}). \label{fig:O32_vs_R3_and_R23}}
\end{figure*}

The chemical properties of galaxies provide powerful constraints on models of galaxy formation and evolution. For example, the abundance of heavy elements (i.e., the metallicity) in the interstellar medium provides crucial information about a galaxy's integrated SFH along with the interplay of outflowing metals and inflowing pristine gas. Furthermore, each heavy element has a different formation pathway with a characteristic timescale and, thus, their relative abundances probe the history of stellar mass assembly and star formation of a galaxy. JWST has revealed a complex picture of chemical abundance patterns in galaxies at the redshift frontier, including the discovery of galaxies with diverse, nonsolar abundance patterns. Currently, it is unclear what exactly is causing the diversity in these nonsolar abundance patterns, although proposed explanations have included enrichment from Population III stars and protoglobular cluster formation in which runaway stellar collisions could produce remarkable objects, such as supermassive stars.

\begin{figure}[t!]
    \centering
    \includegraphics[width=1.0\linewidth]{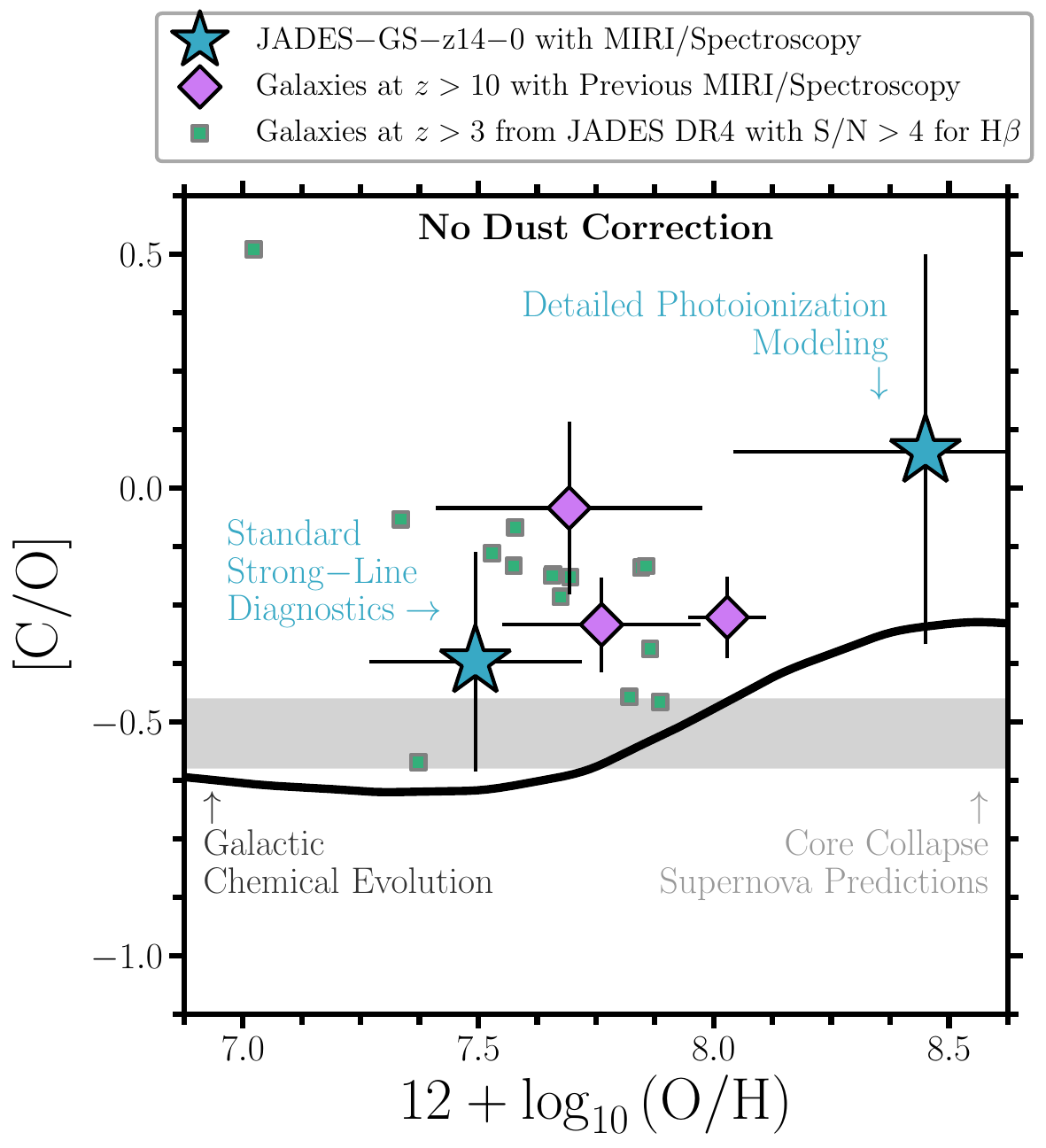}
    \caption{\textbf{Carbon-to-oxygen ratio versus gas-phase oxygen abundance.} The inferred chemical abundance patterns for JADES-GS-z14-0 are compared with the same three samples of star-forming galaxies at $z > 3$ from Figure~\ref{fig:LHalpha_vs_MUV}. There are two measurements provided for JADES-GS-z14-0; one at low-metallicity from standard strong-line diagnostics (Section~\ref{SectionFiveFour}) and the other at high-metallicity from detailed photoionization modeling (Section~\ref{SectionSix}). No dust corrections have been applied to any of these galaxies. With the strong-line diagnostics, the inferred abundance patterns are highly sensitive to the assumed electron temperature. We also compare with the predicted yields from core-collapse supernovae \citep[][]{Tominaga:2007} as illustrated by the gray shaded region and those from galactic chemical evolution models \citep[][]{Kobayashi:2020} as illustrated by the solid black line. The inferred chemical abundance pattern for \mbox{JADES-GS-z14-0} is consistent with predicted yields from core-collapse supernovae for a young, low-metallicity galaxy.\label{fig:logCtoO_vs_logOtoH}}
\end{figure}

A standard way to infer gas-phase oxygen abundances is using the flux ratios of rest-frame optical emission lines \citep[e.g., for a review, see][]{Maiolino:2019}, such as the commonly used $R_{3}$ and $R_{23}$ indices. $R_{3}$ is defined as the flux ratio $\mathrm{[OIII]} \lambda 5007 / \mathrm{H}\beta$ while $R_{23}$ is defined as $\{ \mathrm{[OII]} \lambda\lambda 3727{,}3729 + \mathrm{[OIII]} \lambda\lambda 4959{,}5007 \} / \mathrm{H}\beta$. The standard strong-line diagnostics $O_{32}$ (see also Section~\ref{SectionFiveFive}) versus $R_{3}$ and $R_{23}$ are illustrated in Figure~\ref{fig:O32_vs_R3_and_R23}. This figure also shows brown vertical lines of constant metallicity and brown horizontal lines of constant ionization parameter. We use Equation~\ref{eq:metallicity_calibrations} to determine metallicities, where $Z$ refers to the inferred gas-phase oxygen abundance $12 + \mathrm{log}_{10} (\mathrm{O/H}) - 8.69$, $C_{i}$ are coefficients for the adopted metallicity calibration from Appendix~B of \citet[][]{Curti:2024}, and $R$ is the relevant flux ratio. \begin{equation}
    \label{eq:metallicity_calibrations}
    \mathrm{log}_{10} (R) = \sum_{i=0}^{n} C_{i} Z^{i},
\end{equation}


Using the $R_{3}$ and $R_{23}$ indices, we derive a gas-phase oxygen abundance of $\mathrm{log}_{10} (\mathrm{O/H}) = -1.2 \pm 0.2$ for JADES-GS-z14-0 (which corresponds to $\approx 6\%\ Z_{\odot}$ with a $1 \sigma$ confidence interval of $\approx 4-10\%\ Z_{\odot}$).

There is a well-known, double-valued degeneracy between the gas-phase oxygen abundance and our adopted strong-line diagnostics ($R_3$ and $R_{23}$). The turnover in the relation between abundance and emission line ratios occurs at $\approx 25\%\ Z_{\odot}$, where lower values correspond to lower stellar masses and higher values correspond to higher stellar masses. In order to convert the measured line ratios into oxygen abundances, we assume the lower values to break the double-valued degeneracy. This is one of many limitations in using standard strong-line diagnostics to infer physical properties for the most distant galaxies. In Section~\ref{SectionSeven}, we additionally infer the gas-phase oxygen abundance using detailed photoionization modeling and find much larger values ($\approx 60\%\ Z_{\odot}$). This is consistent with results from strong-line diagnostics if we were to instead assume the higher values to break the double-valued degeneracy in $R_3$ and $R_{23}$.

Following the methodology from \citet[][]{Hsiao:2025}, we further infer the carbon-to-oxygen ratio $\mathrm{C/O}$ using the $\mathrm{CIII]} \lambda\lambda 1907{,}1909$ flux from \citet[][]{Carniani:2024}. To accomplish this, we calculate the doubly ionized carbon-to-oxygen ratio $\mathrm{C^{++}/O^{++}}$ and then correct for unobserved ionic species using an ionization correction factor from \citet[][]{Berg:2019}. Assuming a gas-phase metallicity of $10\%\ Z_{\odot}$, we use the $3\sigma$ lower limit on the ionization parameter from Section~\ref{SectionFiveFive} to determine an ionization correction factor of $\mathrm{ICF} > 0.98$ (since $\mathrm{[OII]} \lambda\lambda 3727{,}3729$ is undetected). We assume that $\mathrm{CIII]} \lambda\lambda 1907{,}1909$ and $\mathrm{[OIII]} \lambda\lambda 4959{,}5007$ arise from nebular gas with the same electron temperature and density. Using these assumptions, the inferred ionization correction factor, and the getIonAbundance function from the \texttt{PyNeb} package \citep[][]{Luridiana:2015}, we determine a carbon-to-oxygen ratio of $\mathrm{log}_{10} (\mathrm{C/O}) = -0.37 \pm 0.24$ for \mbox{JADES-GS-z14-0}. The inferred abundance pattern is largely insensitive to the assumed density. However, the inferred carbon-to-oxygen ratios are $\approx 3-5 \times$ smaller if we assume a temperature of $1.5 \times 10^{4}\ \mathrm{K}$ instead of $10^{4}\ \mathrm{K}$. Figure~\ref{fig:logCtoO_vs_logOtoH} shows the carbon-to-oxygen ratio versus gas-phase oxygen abundance for \mbox{JADES-GS-z14-0} alongside some other high-redshift galaxies that have coverage of $\mathrm{CIII]} \lambda\lambda 1907{,}1909$ and $\mathrm{[OIII]} \lambda\lambda 4959{,}5007$. For simplicity and self-consistency, no dust corrections have been applied to any of the galaxies in Figure~\ref{fig:logCtoO_vs_logOtoH}. The inferred chemical abundance pattern for \mbox{JADES-GS-z14-0} is consistent with the predicted yields from core-collapse supernovae for a young, low-metallicity galaxy.

\subsection{Ionization Parameter}
\label{SectionFiveFive}

The vast majority of the commonly observed emission lines arise from ionized atoms, either through recombination emission, such as those from the Balmer hydrogen lines (e.g., $\mathrm{H}\alpha$), or collisionally excited emission, such as those from the ionized oxygen lines (e.g., $\mathrm{[OIII]}$). The absolute and relative strengths of these lines are sensitive to the intrinsic ionizing spectrum and the dominant gas cooling mechanisms (i.e., radiative cooling from ionized metallic emission lines). Thus, the ionization state of gas permeating the interstellar medium is a crucial property to understand about distant galaxies. The ionization state of nebular gas is commonly parametrized by the dimensionless ionization parameter ($U$), which is defined as the ratio between the number density of incident hydrogen-ionizing photons ($q$) and the number density of hydrogen ($n_{\mathrm{H}}$), divided by the speed of light.

It has been shown that the ionization parameter evolves with respect to galaxy properties and redshift \citep[e.g.,][]{Sanders:2016, Strom:2017, Strom:2018}. For example, ionization parameter tends to increase with decreasing stellar mass and gas-phase metallicity along with increasing redshift. The physical explanation for the redshift evolution at fixed stellar mass is that high-redshift galaxies have lower metallicities and higher specific star formation rates than their lower redshift counterparts, which together increase the number of incident hydrogen-ionizing photons. So far, JWST observations have confirmed this redshift evolution into the redshift frontier, finding galaxies with extreme ionization conditions described by ionization parameters that are more than an order of magnitude larger than those of typical star-forming galaxies at lower redshifts \citep[e.g.,][]{Bunker:2023, Calabro:2024, Naidu:2026}. \begin{equation}
    \label{eq:ionization_parameter}
    \mathrm{log}_{10} (U) = 0.79 \times \mathrm{log}_{10} (O_{32}) - 2.95
\end{equation}

As described in Equation~\ref{eq:ionization_parameter}, the standard way to infer the ionization parameter is using the $O_{32}$ index \citep[e.g., for a review, see][]{Kewley:2019}. $O_{32}$ is defined as the flux ratio $\mathrm{[OIII]} \lambda\lambda 4959{,}5007 / \mathrm{[OII]} \lambda\lambda 3727{,}3729$. This equation presents the calibration from \citet[][]{Strom:2018} using a sample of $N \approx 150$ star-forming galaxies at $z \approx 2-3$. A more recent calibration from \citet[][]{Papovich:2022}, which used a sample of star-forming galaxies at $z \approx 1-2$, agrees with our adopted calibration. Since $\mathrm{[OII]} \lambda\lambda 3727{,}3729$ is not well detected, we derive a $3 \sigma$ lower limit of \ResultLogU\ for JADES-GS-z14-0. 

\subsection{Electron Density}
\label{SectionFiveSix}

Many of the aforementioned physical properties have higher-order dependencies on the electron density. At high densities, recombination rates increase and collisional de-excitation begins, at least for certain emission lines with sufficiently low critical densities. For example, the electron density is partially responsible for determining the slope of the mass-metallicity relation and, at high densities, $O_{32}$ begins to trace density more than ionization parameter due to collisional de-excitation of the $[\mathrm{OII}] \lambda\lambda 3727{,}3729$ doublet \citep[e.g.,][]{Choustikov:2026}.

The physical picture of distant galaxies is further complicated by the multiphased structure of the interstellar medium, in which different diagnostics of the electron density suggest entirely different densities. For example, at high redshifts, rest-frame UV tracers for the electron density such as $\mathrm{CIII]} \lambda\lambda 1907{,}1909$ typically imply higher densities than optical tracers such as $[\mathrm{OII}] \lambda\lambda 3727{,}3729$ or $[\mathrm{SII}] \lambda\lambda 6717{,}6731$, which typically imply higher densities than far-infrared tracers such as $[\mathrm{OIII}] \lambda\lambda 52{,}88 \mu\mathrm{m}$ \citep[e.g.,][]{Harikane:2025b}. Furthermore, prior work shows that electron density tends to increase with redshift due to smaller physical sizes and thus larger surface densities \citep[e.g.,][]{Isobe:2023, Topping:2025b}.

For JADES-GS-z14-0, we only have access to one tracer for the electron density due to the low spectral resolution of the available data at rest-frame UV and optical wavelengths. The one diagnostic available to us is the flux ratio of $[\mathrm{OIII}] \lambda 5007$ and $[\mathrm{OIII}] \lambda 88 \mu\mathrm{m}$. To first order, this ratio is sensitive to the electron density, with a secondary dependence on the electron temperature due to significantly different energy levels for the two relevant atomic transitions. There is no dependence on metallicity nor ionization parameter since these two emission lines originate from the same ionic species.

We determine predictions for the relevant emission line ratio of $[\mathrm{OIII}]$ using the getEmissivity function from the \texttt{PyNeb} package \citep[][]{Luridiana:2015}. \mbox{By assuming} an electron temperature of $T_{e} = 10^{4}\ \mathrm{K}$, we first derive a density of $n_{\mathrm{H}} = 690 \pm 200\ \mathrm{cm}^{-3}$ for \mbox{JADES-GS-z14-0}. If we instead assume a temperature of $T_{e} = 1.5 \times 10^{4}\ \mathrm{K}$, we instead derive $n_{\mathrm{H}} = 150 \pm 110\ \mathrm{cm}^{-3}$. Additionally, we explore the predicted $[\mathrm{OIII}]$ ratio across a wide range of electron density ($\mathrm{log}_{10} [n_{\mathrm{H}} / \mathrm{cm}^{-3}] = 0-10$) and temperature ($\mathrm{log}_{10} [T_{e} / \mathrm{K}] = 3.7-4.5$), finding larger (smaller) densities at smaller (larger) temperatures. The adopted parameter ranges are identical to those presented in \citet[][]{Zavala:2025} for GHZ2/GLASSz12.

\section{Inferring Physical Properties Using Detailed Photoionization Modeling}
\label{SectionSix}

\begin{figure*}
    \centering
    \includegraphics[width=1.0\linewidth]{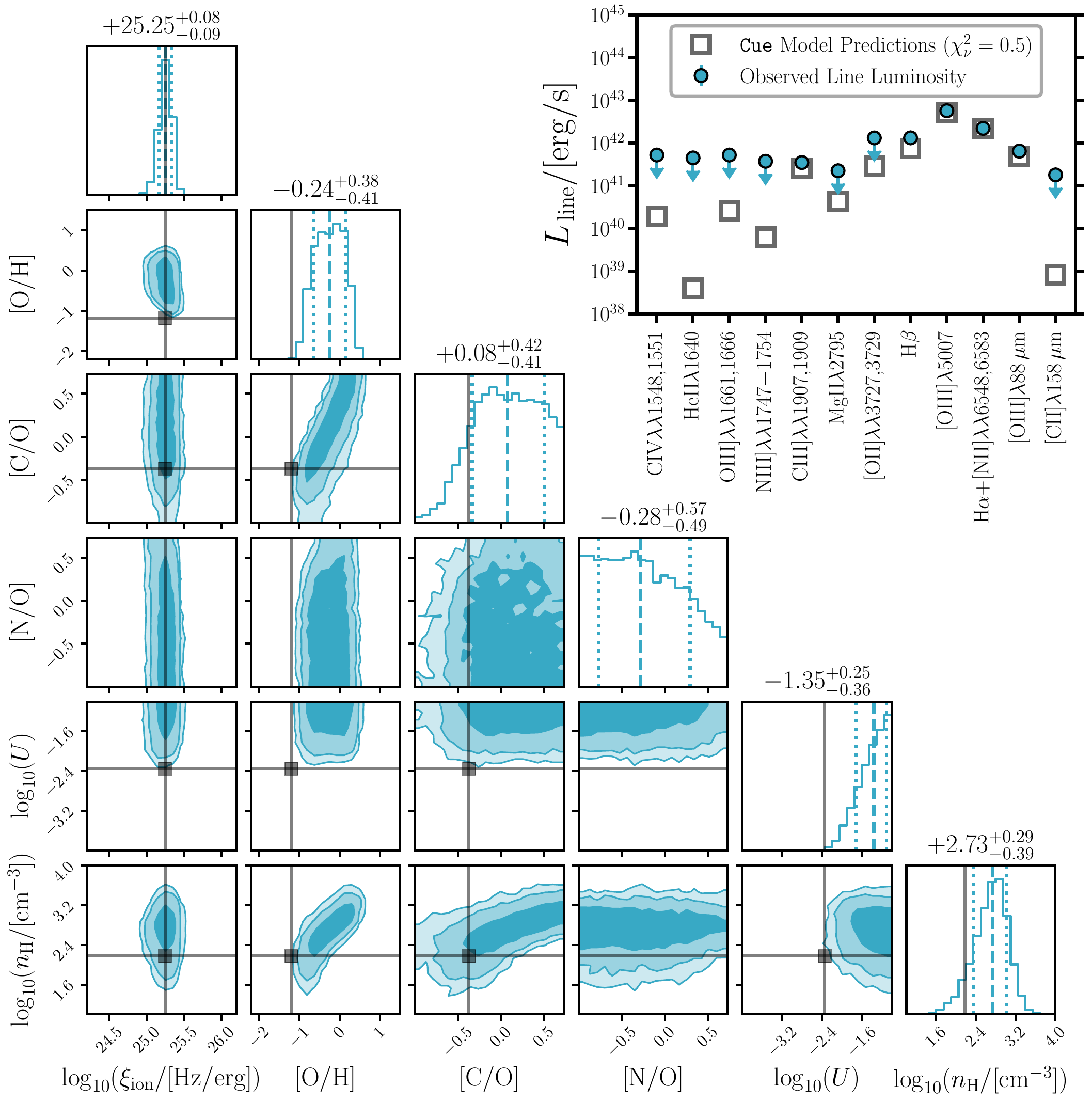}
    \caption{\textbf{Constraints on physical properties using photoionization modeling.} We utilize \texttt{Cue} \citep[][]{Li:2024, Li:2025} to simultaneously infer various physical properties for JADES-GS-z14-0 by modeling observations of nebular emission lines. \textit{Lower left:} The joint posterior distributions are shown for ionizing photon production efficiency, gas-phase oxygen abundance, carbon-to-oxygen ratio, nitrogen-to-oxygen ratio, ionization parameter, and electron density. The contours in the off-diagonal panels represent the $68\%$, $95\%$, and $99\%$ confidence intervals of the modeling. The dashed and dotted lines in the diagonal panels represent the  medians and $68\%$ confidence intervals. The black solid lines represent the properties inferred using the standard strong-line diagnostics described in Section~\ref{SectionFive}. \textit{Upper right:} The predicted line luminosities are compared with observed quantities, demonstrating consistency between the model predictions and observations. \textit{All panels:} The source of the intrinsic ionizing spectrum ($\xi_{\mathrm{ion}}$) along with the state of the interstellar medium ($U$ and $n_{\mathrm{H}}$) are well constrained by existing spectroscopic observations while the chemical abundance patterns ($\mathrm{N/O}$ and $\mathrm{C/O}$) are poorly constrained by the data. The results from our emission line modeling demonstrate extreme ionization conditions and relatively high metallicities within JADES-GS-z14-0. \label{fig:Cue_results_corner}}
\end{figure*}

In Section~\ref{SectionFive}, we explored the physical properties of JADES-GS-z14-0 as inferred from strong-line diagnostics using the brightest rest-frame optical emission lines. This methodology has been used extensively in the literature for all types of galaxies across nearly the full range of cosmic history \mbox{\citep[for a review, see][]{Kewley:2019}}. However, the method of strong-line diagnostics relies on assumptions about unknown physical properties of individual galaxies. For example, by inferring gas-phase metallicities using the $R_{3}$, $R_{23}$, or $\hat{R}$ indices, we implicitly assume that differences in line ratios correspond to differences in an individual physical property -- in this case, gas-phase oxygen abundance. But it is well known that these line ratios are also dependent on many other properties of the interstellar medium, including the excitation and ionization states of the nebular gas. One result of this implicit assumption is that the quoted uncertainties using strong-line diagnostics do not fully encapsulate the uncertainties from the unknown nebular properties of individual galaxies. Thus, the quoted uncertainties for the physical properties inferred in Section~\ref{SectionFive} are likely underestimates of the true values.

To address these issues, we utilize \texttt{Cue} \citep[][]{Li:2024, Li:2025} in order to infer the same physical properties from Section~\ref{SectionFive}, but this time fully accounting for the covariance between each of the inferred properties by constraining all of these properties simultaneously. \texttt{Cue} is a fast and flexible neural network emulator trained on the \texttt{Cloudy} photoionization modeling code \citep[][]{Ferland:1998, Chatzikos:2023}. Importantly, unlike traditional modeling of nebular emission lines, \texttt{Cue} does not require a specific ionizing spectrum as input; rather, it approximates the intrinsic ionizing spectrum with a four-part piecewise power law. This parameterization of the ionizing spectrum is agnostic about the source of the ionizing photons. Thus, \texttt{Cue} is capable of modeling ionizing sources from stellar populations, accreting supermassive black holes, and, perhaps most intriguingly, efficient and/or hard ionizing sources that are outside of our known modeling space. \texttt{Cue} has already been used to infer the properties of the most distant galaxy currently known, MoM-z14 at $z = 14.44$ \citep[][]{Naidu:2026}.

Figure~\ref{fig:Cue_results_corner} provides constraints on the joint posterior distributions for some of the most important physical properties inferred using the strong-line diagnostics described in Section~\ref{SectionFive}. In the lower left, from left to right, these properties include the ionizing photon production efficiency ($\xi_{\mathrm{ion}}$), gas-phase oxygen abundance ($\mathrm{[O/H]}$), carbon-to-oxygen ratio ($\mathrm{[C/O]}$), nitrogen-to-oxygen ratio ($\mathrm{[N/O]}$), ionization parameter ($U$), and electron density ($n_{\mathrm{H}}$). The dashed lines represent the medians of the \texttt{Cue} predictions while the dotted lines represent the $68\%$ confidence intervals. In the upper right, we show the comparison between the observed and predicted fluxes for each of the nebular emission lines that were used as input into \texttt{Cue}. Note that we only have upper limits for some of the provided emission lines and we include these limits because they are still informative. The \texttt{Cue} models do well at reproducing the observations, as demonstrated by the reduced chi-squared statistic, which is less than one. This figure highlights constraints for each of the inferred physical properties, but also degeneracies between some of these properties. Most notably, there is a degeneracy between the gas-phase oxygen abundance and electron density, in which we infer smaller metallicities for smaller densities and higher metallicities for higher densities. We should also note that the nitrogen-to-oxygen ratio is almost entirely unconstrained since none of the nitrogen lines are detected, although the upper limit on the $\mathrm{NIII]} \lambda\lambda 1747{-}1754$ quintuplet provides a small, nonzero constraint on this value.

To quickly summarize these results from the detailed photoionization modeling using \texttt{Cue}, we generally find consistency between the physical properties inferred \mbox{using} this method and those same properties inferred using the standard strong-line diagnostics described in Section~\ref{SectionFive}. The most notable difference between these methods is in the inferred gas-phase oxygen abundance, for which we derive $\approx 6\%\ Z_{\odot}$ when using the standard strong-line diagnostics, but $\approx 60\%\ Z_{\odot}$ when using \texttt{Cue}'s detailed photoionization modeling. Another notable difference is in the inferred uncertainties, which are $2-5 \times$ larger when using \texttt{Cue}'s photoionization modeling that properly accounts for the covariances between each of the inferred properties. \texttt{Cue} predicts that $\mathrm{[OIII]} \lambda 52\mu\mathrm{m}$ should be one of the brightest emission lines for \mbox{JADES-GS-z14-0}, roughly $2-3 \times$ stronger than $\mathrm{[OIII]} \lambda 88\mu\mathrm{m}$. It is important to note that the relative strength of $\mathrm{[OIII]} \lambda 52\mu\mathrm{m}$ and $\mathrm{[OIII]} \lambda 88\mu\mathrm{m}$ is highly sensitive to the electron density. Since there is a degeneracy between the gas-phase oxygen abundance and electron density, we would recommend obtaining observations of $\mathrm{[OIII]} \lambda 52\mu\mathrm{m}$ for JADES-GS-z14-0 to improve constraints on these degenerate physical properties. However, $\mathrm{[OIII]} \lambda 52\mu\mathrm{m}$ for JADES-GS-z14-0 directly coincides with an atmospheric absorption feature and is thus unobservable with ground-based observatories.

We performed one more iteration of photoionization modeling using \texttt{Cue}, but this time without fitting the rest-frame optical emission lines in order to explore the impact of including those lines. This informs us about the amount of information that we gained after acquiring the ultra-deep, low-resolution, mid-infrared spectroscopy of JADES-GS-z14-0 presented in this work. The most notable difference is in the predicted ionizing photon production efficiency, for which we infer values of $\mathrm{log}_{10} (\xi_{\mathrm{ion}} / [\mathrm{Hz/erg}]) = 25.25_{-0.09}^{+0.08}$ with the full dataset and $\mathrm{log}_{10} (\xi_{\mathrm{ion}} / [\mathrm{Hz/erg}]) = 25.23_{-0.34}^{+0.43}$ without fitting the rest-frame optical emission lines. Similarly, the other notable difference is in the electron density, for which we infer values of $\mathrm{log}_{10} (n_{\mathrm{H}} / [\mathrm{cm}^{-3}]) = 2.73_{-0.39}^{+0.29}$ with the full dataset and $\mathrm{log}_{10} (n_{\mathrm{H}} / [\mathrm{cm}^{-3}]) = 2.27_{-0.65}^{+0.69}$ without fitting the rest-frame optical emission lines. The remainder of the inferred properties and their associated uncertainties are identical between the two iterations of \texttt{Cue}'s photoionization modeling. Thus, the ultra-deep mid-infrared spectroscopy that we acquired resulted in ionizing photon production efficiencies and electron densities that are $2-3 \times$ better constrained, highlighting the importance of obtaining mid-infrared observations for understanding galaxies at the redshift frontier.

\section{Results, Interpretation, \& Discussion}
\label{SectionSeven}

\subsection{Trying to Explain the Surprising Abundance of Luminous Galaxies in the Early Universe}
\label{SectionSevenOne}

Despite remarkable progress with identifying and understanding galaxies in the early Universe, JWST has revealed a profound puzzle -- galaxies at the redshift frontier are dramatically more numerous and luminous than predicted by many theoretical models produced prior to the launch of JWST. Several competing, and possibly complementary, hypotheses have emerged in the last few years to explain these surprising observations:
\begin{enumerate}[label=\textbf{\#\arabic*}]
    \itemsep 0pt
    \item \ul{\textit{Bursty SFHs}}: High-redshift galaxies have been predicted to experience intense episodic bursts of star formation \citep[e.g.,][]{Mason:2023, Shen:2023, Sun:2023}. During these bursts, the youngest and most massive stars will dominate the light and ``outshine'' the older stellar population that dominates the mass. Thus, galaxies will appear more luminous during bursts of star formation, while obscuring evidence of prior and extended star formation. To put it another way, bursty SFHs would increase the scatter in the relation between UV luminosity ($L_{\mathrm{UV}}$) and halo mass ($M_{\mathrm{halo}}$). However, the timescales and physical origins of these bursts remain poorly constrained.
    \item \ul{\textit{Increased SFEs}}: Galaxies in the early Universe may have been able to convert nebular gas into stars much more efficiently \citep[e.g.,][]{Dekel:2023, Feldmann:2025, Jeong:2025, Mauerhofer:2025}. This increased efficiency is possibly caused by feedback-free starbursts (FFBs), when the high densities and low metallicities at high redshifts guarantee increased SFEs in the most massive dark matter halos due to ineffective feedback processes when compared to the local Universe. FFBs occur when the free-fall time is shorter than the time it takes for low-metallicity stars to develop winds and produce supernovae. However, the magnitude and physical origins of these efficiencies are not fully understood.
    \item \ul{\textit{Different IMFs}}: The first galaxies are expected to form stars dissimilarly to galaxies in the local Universe \citep[e.g.,][]{Cueto:2024, Trinca:2024, Yung:2024, Jeong:2025, Liu:2025, Mauerhofer:2025}. For example, it is likely that the formation of low-mass stars is suppressed (``bottom-light'' IMF) and the formation of high-mass stars is enhanced (``top-heavy'' IMF) in the early Universe due to lower metallicities and an increased temperature of the cosmic microwave background ($T \approx 60\ \mathrm{K}$ at $z = 20$). These modifications to the IMF would produce brighter galaxies for a given amount of stellar mass because massive stars are more luminous than lower mass stars. These modifications would also accelerate metal enrichment due to increased supernova rates. If the IMF extends to stars with masses larger than $200\ M_{\odot}$, pair-instability supernovae would further enhance metal yields \citep[][]{Liu:2025}. However, there is still no convincing evidence for IMF variations in the early Universe.
\end{enumerate}

Thankfully, these three competing hypotheses have distinct predictions for the properties of the most distant galaxies \citep[e.g.,][]{Choustikov:2026}. Let us present the evidence that we have assembled for \mbox{JADES-GS-z14-0}. The measured rest-frame optical equivalent widths for \mbox{JADES-GS-z14-0} are consistent with but smaller than those measured for similarly luminous galaxies at lower redshifts \citep[$z \approx 8$;][]{Endsley:2024}. Additionally, the shorter timescale star formation rate inferred from $\mathrm{H}\alpha$ is much smaller than the longer timescale star formation rate inferred from rest-frame UV continuum emission (see the discussion at the end of Section~\ref{SectionFiveTwo}). To put it another way, the luminosity of $\mathrm{H}\alpha$ (i.e., $L_{\mathrm{H}\alpha}$) divided by that of the UV continuum (i.e., $L_{\mathrm{UV}}$) suggests ionizing photon production efficiencies similar to the canonical value (i.e., $\xi_{\mathrm{ion}} \approx 10^{25.2}\ \mathrm{Hz/erg}$) and far below the stellar population maximum (i.e., $\xi_{\mathrm{ion}} \approx 10^{26.0}\ \mathrm{Hz/erg}$; see Figure~\ref{fig:xi_ion_vs_zSpec_and_MUV} and the discussion within Section~\ref{SectionFiveThree}).

Using these empirical and inferred properties, we can rule out top-heavy IMFs for JADES-GS-z14-0, since its ionizing photon production efficiencies are far below the stellar population maximum. At first glance, it appears that we can rule out bursty SFHs because the equivalent widths measured for JADES-GS-z14-0 are not extreme. However, a lull (or decline) in the most recent few million years of the SFH would naturally explain the discrepancy between star formation rates inferred from the shorter timescale emission lines and the longer timescale rest-frame UV continuum. For these reasons, out of the three hypotheses presented here, we argue that increased SFEs and bursty SFHs are the two most likely explanations for the existence of JADES-GS-z14-0, although we cannot definitively rule out different IMFs, especially those with fewer low-mass stars (i.e., bottom-light).

\subsection{JADES-GS-z14-0 in the Broader Context of Luminous Galaxies at $\mathit{z > 10}$}
\label{SectionSevenTwo}

\citet[][]{Harikane:2025a} and \citet[][]{Naidu:2026} found evidence for a size-chemistry bimodality in the population of luminous ($M_{\mathrm{UV}} < -20$) galaxies at $z > 10$ with spectroscopic confirmation. These works showed that extended high-redshift sources tend to be nitrogen-weak with relatively weak \mbox{rest-frame} UV emission lines (e.g., JADES-GS-z14-0) while compact high-redshift sources tend to be nitrogen-strong with relatively strong UV lines (e.g., GN-z11, GHZ2/GLASSz12, and MoM-z14). At first glance, these differences in apparent sizes, chemical abundance patterns, and line strengths suggest the existence of at least two fundamentally different formation pathways among luminous galaxies at $z > 10$. But are two formation pathways truly necessary to explain the population of luminous galaxies at $z > 10$?

\citet[][]{Roberts-Borsani:2026} provide an alternative interpretation for these two populations of galaxies at $z > 10$, suggesting they are short-lived snapshots along a common evolutionary pathway; a similar interpretation is provided by \citet[][]{Tang:2026}. Their arguments are that compact high-redshift sources with relatively strong rest-frame UV lines are experiencing a recent burst in the last few million years of their SFHs while extended high-redshift sources with relatively weak UV lines are experiencing a recent lull in their SFHs. Additionally, \citet[][]{Tang:2026} find that the prevalence of galaxies with bursty SFHs increases with redshift.

Our results are consistent with the interpretation from \citet[][]{Roberts-Borsani:2026} and \citet[][]{Tang:2026}. For the four luminous galaxies at $z > 10$ that have MIRI/Spectroscopy, we find similar optical emission line ratios ($R_{3}$, $R_{23}$, and $O_{32}$), indicating similar conditions in the interstellar mediums of these galaxies. However, despite their similar luminosities, \mbox{JADES-GS-z14-0} and \mbox{MACS0647-JD1} have smaller ionizing photon production efficiencies ($\xi_{\mathrm{ion}} \approx 10^{25.3}\ \mathrm{Hz/erg}$) when compared to GN-z11 and GHZ2/GLASSz12 ($\xi_{\mathrm{ion}} \approx 10^{25.6}\ \mathrm{Hz/erg}$). A recent lull (or decline) in the SFH would naturally explain this difference. Furthermore, \mbox{JADES-GS-z14-0} has a larger physical size ($r_{\mathrm{UV}} \approx 260\ \mathrm{pc}$) and thus a smaller stellar surface density when compared to \mbox{GN-z11} and \mbox{GHZ2/GLASSz12} ($r_{\mathrm{UV}} \approx 40-60\ \mathrm{pc}$), while \mbox{MACS0647-JD1} has multiple components and possibly represents an ongoing galaxy merger. These differences could possibly be explained by inside-out growth, in which compact sources are forming their cores while extended sources are forming their disks.

We additionally compare JADES-GS-z14-0's inferred physical properties from \texttt{Cue} with those same properties inferred for MoM-z14 at $z > 14$ \citep[][]{Naidu:2026}. The modeling assumptions are identical and, thus, this represents close to a one-to-one comparison between the two most distant galaxies currently known, modulo differences in available data. The most important difference is the additional constraints for some of the strongest rest-frame optical and far-infrared emission lines for \mbox{JADES-GS-z14-0}, which are absent for \mbox{MoM-z14}; there are at least twice as many observables used to constrain the photoionization model for \mbox{JADES-GS-z14-0}. When compared to MoM-z14, we find that \mbox{JADES-GS-z14-0} has a notably less efficient ionizing spectrum, larger gas-phase oxygen abundance, larger carbon-to-oxygen ratio, smaller nitrogen-to-oxygen ratio, similar ionization parameter, and smaller electron density. Empirically, these differences are best demonstrated by the rest-frame UV emission lines, which are $2-5 \times$ stronger for MoM-z14 than JADES-GS-z14-0 \citep[][]{Carniani:2024}. The extreme ionizing photon production efficiency of \mbox{MoM-z14} ($\xi_{\mathrm{ion}} \approx 10^{26.3 \pm 0.5}\ \mathrm{Hz/erg}$) is $\approx 3-5 \times$ higher than \mbox{GHZ2/GLASSz12} \citep[][]{Zavala:2025} and even \mbox{GN-z11} \citep[][]{Alvarez-Marquez:2025}, making it unique among luminous galaxies at $z > 10$. Spectroscopic observations with JWST/MIRI are necessary to confirm the extreme ionizing photon production efficiency of \mbox{MoM-z14}.

\subsection{Comparison with Theoretical Predictions}
\label{SectionSevenThree}

JADES-GS-z14-0 is unlike theoretical predictions for galaxies at $z > 14$ because of its extreme redshift and combination of inferred stellar mass ($M_{\ast} \gtrsim 10^{8}\ M_{\odot}$) and gas-phase metallicity ($Z_{\mathrm{gas}} \gtrsim 10\%\ Z_{\odot}$). Most recently, \citet[][]{Kim:2026} performed cosmological zoom-in simulations using six different state-of-the-art galaxy formation simulation codes (ENZO, RAMSES, CHANGA, GADGET-3, GADGET-4, and GIZMO). They focused on studying high-redshift galaxies ($z > 10$) that have dark matter halo masses with $10^{10}-10^{11}\ M_{\odot}$ at $z = 10$. Their goal is to test the performance of galaxy formation models developed for lower redshifts without having to invoke additional physics for the high-redshift environments. We find that their simulations are unable to reproduce any of the empirical and inferred properties for JADES-GS-z14-0, including absolute UV magnitude, stellar mass, star formation rate, and gas-phase metallicity. At $z > 14$, the most luminous and massive galaxy in their simulations has $M_{\mathrm{UV}} \approx -18$, $M_{\ast} \approx 10^{7}\ M_{\odot}$, $\mathrm{SFR} \approx 1\ M_{\odot} / \mathrm{yr}$, and $Z_{\mathrm{gas}} \approx 1\%\ Z_{\odot}$. Each of these values is at least an order of magnitude smaller than the same empirical and inferred properties for \mbox{JADES-GS-z14-0}. Even at $z = 10$, there are zero galaxies that are similar to \mbox{JADES-GS-z14-0}. \citet[][]{Kim:2026} conclude that the properties of high-redshift galaxies are extremely sensitive to assumptions about feedback. Increasing the SFE of the simulations by an order of magnitude, from a few percent to a few tens of percent, would potentially rectify these differences between observations and theoretical predictions.

We additionally compare with other large-scale, state-of-the-art simulations of galaxy formation, including the First Light And Reionization Epoch Simulations \citep[FLARES;][]{Wilkins:2023}, the MillenniumTNG project \citep[][]{Kannan:2023}, and the THESAN-ZOOM project \citep[][]{Kannan:2025}. These are some of the only galaxy simulations with comparable predictions at $z > 10$. They each adopt different resolutions, feedback prescriptions, and assumptions about the multiphased interstellar medium. Yet none of these simulations are able to reproduce the empirical and inferred properties of JADES-GS-z14-0, supporting our conclusion that this distant galaxy is unlike theoretical predictions for galaxy formation and evolution in the early Universe.

\section{Summary \& Conclusions}
\label{SectionEight}

JADES-GS-z14-0 is challenging our understanding of the early Universe as one of the archetypes for the early period of galaxy formation and evolution that has been discovered by JWST. In this work, we provided a first look at ultra-deep ($t_{\mathrm{obs}} \approx 51\ \mathrm{hr}$) spectroscopic follow-up observations of \mbox{JADES-GS-z14-0} with JWST/MIRI's Low Resolution Spectrometer (LRS), which is by far the longest spectroscopic integration ever acquired by JWST/MIRI. These recently acquired observations cover the brightest rest-frame optical nebular emission lines ($[\mathrm{OII}] \lambda\lambda 3727{,}3729$, $\mathrm{H}\beta$, $[\mathrm{OIII}] \lambda\lambda 4959{,}5007$, and $\mathrm{H}\alpha$) in addition to the stellar continuum for one of the most distant galaxies currently known at $z = 14.18$. By studying these prominent emission lines, we unveiled important insights into the properties of this extreme galaxy, less than $300$ million years after the Big Bang. Our findings can be summarized as follows.
\begin{itemize}
    \itemsep 0pt
    \item We clearly detect two features above the noise level in the 2D and 1D spectra (see Figure~\ref{fig:Full_Spectrum}) along with the continuum. These spectroscopic features include $[\mathrm{OIII}] \lambda\lambda 4959{,}5007$ detected at $\approx 14\sigma$ and $\mathrm{H}\alpha$ detected at $\approx 4\sigma$ (see also Figure~\ref{fig:ZoomIn_EmissionLines}), which are the most distant detections of these lines. We note that $[\mathrm{OII}] \lambda\lambda 3727{,}3729$ is undetected while $\mathrm{H}\beta$ is only marginally detected at $\approx 3 \sigma$.
    \item We directly determine the Balmer decrement and find \ResultBD. This is fully consistent with Case~B recombination and thus no measurable dust attenuation in \mbox{JADES-GS-z14-0}.
    \item We directly measure the $\mathrm{H}\alpha$ line luminosity (see Figure~\ref{fig:LHalpha_vs_MUV}) and infer a recent star formation rate of \ResultSFR. The inferred star formation rate surface density (\ResultSigmaSFR) is consistent with values measured for similarly luminous galaxies at $z \approx 8$ and comparable to the most vigorous starbursts observed at $z \approx 0$.
    \item By comparing the luminosities of $\mathrm{H}\alpha$ and the rest-frame UV continuum, we infer an ionizing photon production efficiency of \ResultXiIon\ (see Figure~\ref{fig:xi_ion_vs_zSpec_and_MUV}). Our inference is consistent with the canonical value observed locally, but significantly lower than the stellar population maximum along with values inferred for similarly luminous galaxies at $z > 10$, such as \mbox{GN-z11} and \mbox{GHZ2/GLASSz12}.
    \item By comparing the strengths of collisionally excited oxygen lines and recombination hydrogen lines, we measure the $R_{3}$ and $R_{23}$ indices (see Figure~\ref{fig:O32_vs_R3_and_R23}) then use these to infer the gas-phase oxygen abundance, finding \ResultOtoH. The inferred gas-phase metallicity is consistent with $\approx 6\%\ Z_{\odot}$ and similar to values measured for galaxies at $z > 3$ from JADES DR4. Detailed photoionization modeling suggests much larger metallicities of $\approx 60\%\ Z_{\odot}$ (see Figure~\ref{fig:Cue_results_corner}), although there appears to be a strong degeneracy between metallicity and density. We further infer a carbon-to-oxygen ratio of \ResultCtoO\ using $\mathrm{CIII]} \lambda\lambda 1907{,}1909$.
    \item We provide a limit on the $O_{32}$ index (see Figure~\ref{fig:O32_vs_R3_and_R23}), then use this to infer the ionization parameter and find a $3\sigma$ lower limit of \ResultLogU.
    \item By comparing the luminosities of $[\mathrm{OIII}] \lambda 5007$ and $[\mathrm{OIII}] \lambda 88 \mu\mathrm{m}$, we infer the electron density and find \ResultDensity\ if we assume an electron temperature of $T_{e} = 10^{4}\ \mathrm{K}$. The inferred density is larger than typical conditions seen in the local Universe ($z \approx 0$) and provides further evidence for a redshift evolution toward increasing electron densities at high redshifts. This evolution might be driven by increased star formation rate surface densities ($\Sigma_{\mathrm{SFR}}$) at high redshifts.
\end{itemize}

Altogether, our results suggest that \mbox{JADES-GS-z14-0} is a prolific producer of ionizing photons and a significant fraction of these are possibly escaping the galaxy into the intergalactic medium (IGM). These ionizing photons are creating one of the earliest ionized regions in the IGM, less than $300$ million years after the Big Bang, when the Universe was only $\approx 2\%$ of its current age.

Furthermore, our results confirm the relatively high metal content within \mbox{JADES-GS-z14-0}, suggesting rapid metal enrichment, which existing simulations struggle to reproduce so early in cosmic history (see the discussion in Section~\ref{SectionSevenThree}). The presence of metals indicates that massive stars have already lived, died, and enriched the interstellar material of \mbox{JADES-GS-z14-0} in only a few hundred million years. By studying some of the first generations of stars and the buildup of heavy metals at the break of cosmic dawn, our findings have important implications for the understanding of galaxy formation and evolution in the early Universe. In this work, we focused our attention on the rest-frame optical emission lines of \mbox{JADES-GS-z14-0}, while the rest-frame optical continuum will be discussed and interpreted in a forthcoming manuscript from the JADES collaboration (Helton et~al. 2026, \mbox{in preparation}). This forthcoming work will simultaneously model the properties of the stars, nebular gas, and dust by fitting the panchromatic spectral energy distribution (SED) of \mbox{JADES-GS-z14-0}.

The primary limitation of our work is its focus on a single galaxy at the redshift frontier, since it is \mbox{unclear} if JADES-GS-z14-0 is representative of the high-redshift star-forming galaxy population due to its spatial extent and relatively weak nebular emission lines. Thankfully, upcoming observations will address the primary limitation of our work. JWST Cycle 4 devoted an enormous amount of mission time ($t \approx 330\ \mathrm{hr}$) in order to obtain mid-infrared spectroscopy for a relatively large sample of galaxies at the redshift frontier, including a total of $15$ galaxies at $z \gtrsim 10$ (see proposal IDs: $7078$, $8051$, $8544$, $9165$, $9425$); JWST/MIRI will observe these galaxies in the next year. These observations will expand our work by improving our understanding of the rest-frame optical emission lines for the highest redshift galaxies, since these lines encode valuable information about the properties of the stars, nebular gas, and dust. The manual postprocessing and custom corrections that we developed for this work are specialized for long MIRI/LRS integrations and will ultimately increase the scientific return for many of these upcoming observations.

\section*{Acknowledgments}

We thank the anonymous referee for providing thorough and constructive feedback that helped improve this paper. We thank Greg Sloan and Carol Rodriguez as the instrument scientist reviewer and program coordinator for JWST Program \#8544, respectively, for critical advice when designing and optimizing the observing strategy. Finally, we thank Jorge Zavala, Sarah Kendrew, and Javier \'{A}lvarez-M\'{a}rquez for valuable conversations that ultimately strengthened this paper. This work is based on observations made with the NASA/ESA/CSA James Webb Space Telescope (JWST). These data were obtained from the Mikulski Archive for Space Telescopes (MAST) at the Space Telescope Science Institute, which is operated by the Association of Universities for Research in Astronomy, Inc., under NASA contract \mbox{NAS 5-03127} for JWST. These observations are associated with JWST Program \#8544, which can be accessed using \dataset[DOI: 10.17909/vpjw-b773]{https://doi.org/10.17909/vpjw-b773} \citep{PID_8544_DOI}. This work also made use of the {\it lux} supercomputer at UC Santa Cruz, which is funded by NSF MRI grant AST 1828315, in addition to the High Performance Computing (HPC) resources at the University of Arizona, which is funded by the Office of Research Discovery and Innovation (ORDI), Chief Information Officer (CIO), and University Information Technology Services (UITS).

We respectfully acknowledge that The Pennsylvania State University campuses are located on the original homelands of the Erie, Haudenosaunee (Seneca, Cayuga, Onondaga, Oneida, Mohawk, and Tuscarora), Lenape (Delaware Nation, Delaware Tribe, Stockbridge-Munsee), Monongahela, Shawnee (Absentee, Eastern, and Oklahoma), Susquehannock, and Wahzhazhe (Osage) Nations. As a land grant institution, we acknowledge and honor the traditional caretakers of these lands and strive to understand and model their responsible stewardship. We also acknowledge the longer history of these lands and our place in that history.

We respectfully acknowledge that the University of Arizona is on the land and territories of Indigenous peoples. Today, Arizona is home to 22 federally recognized tribes, with Tucson being home to the O’odham and the Yaqui. The University of Arizona strives to build sustainable relationships with sovereign Native Nations and Indigenous communities through education offerings, partnerships, and community service.

J.M.H. acknowledges support from the Evolving Universe Fellowship, which is made possible by a generous donation from Dr. Keiko Miwa Ross. J.M.H., K.N.H., G.H.R., C.N.A.W., E.E., D.J.E., Z.J., B.D.J., M.J.R., and B.R. acknowledge support from the JWST Near Infrared Camera (NIRCam) \mbox{Science} Team Lead, NAS5-02105, from NASA Goddard Space Flight Center to the University of Arizona. J.M.H., D.J.E., and B.R. acknowledge support from JWST Program \#3215. J.M.H., J.E.M., K.N.H., and G.H.R. acknowledge support from JWST Program \#8544. J.E.M., G.H.R., and S.A. acknowledge support from the JWST Mid-Infrared Instrument (MIRI) Science Team Lead, 80NSSC18K0555, from NASA Goddard Space Flight Center to the University of Arizona. S.C. and G.V. acknowledge support by European Union’s HE ERC Starting grant No.~101040227 (WINGS). W.M.B. acknowledges support from DARK via the DARK fellowship; this work was supported by a research grant (VIL54489) from Villum Fonden. A.J.B., J.C., and A.S. acknowledge funding from the ``FirstGalaxies'' Advanced Grant from the European Research Council (ERC) under the European Union’s Horizon 2020 research and innovation program (grant agreement No. 789056). \mbox{E.C.-L.} acknowledges support of an STFC Webb Fellowship (ST/W001438/1). D.J.E. is also supported as a Simons Investigator. R.M. and J.S. acknowledge support by the Science and Technology Facilities Council (STFC), by the ERC through Advanced Grant 695671 ``QUENCH.'' R.M. also acknowledges funding from the UKRI Frontier Research grant RISEandFALL along with a research professorship through the Royal Society. \mbox{P.G.P.-G.} acknowledges support from grant PID2022-139567NB-I00 funded by Spanish Ministerio de Ciencia e Innovaci\'{o}n MCIN/AEI/10.13039/501100011033, FEDER, UE. S.T. acknowledges support by the Royal Society Research Grant G125142. H.\"{U}. acknowledges funding by the European Union (ERC APEX, 101164796); views and opinions expressed are however those of the authors only and do not necessarily reflect those of the European Union or the ERC Executive Agency; neither the European Union nor the granting authority can be held responsible for them. C.C.W. acknowledges support from NOIRLab, which is managed by the Association of Universities for Research in Astronomy (AURA) under a cooperative agreement with the National Science Foundation. J.W. acknowledges support from the Cosmic Dawn Center through the DAWN Fellowship; the Cosmic Dawn Center (DAWN) is funded by the Danish National Research Foundation under grant No.~140.

\facilities{JWST (MIRI, NIRCam, and NIRSpec), ALMA}

\software{\texttt{AstroPy} \citep[][]{Astropy:2013, Astropy:2018}, \texttt{corner} \citep[][]{Foreman-Mackey:2016}, \texttt{Cue} \citep[][]{Li:2024, Li:2025}, \texttt{dynesty} \citep[][]{Speagle:2020}, \texttt{LMFIT} \citep[][]{Newville:2014}, \texttt{Matplotlib} \citep[][]{Matplotlib:2007}, \texttt{NumPy} \citep[][]{NumPy:2011, NumPy:2020}, \texttt{pandas} \citep[][]{Pandas:2022}, \texttt{photutils} \citep[][]{Bradley:2025}, \texttt{PyNeb} \citep[][]{Luridiana:2015}, \texttt{SciPy} \citep[][]{SciPy:2020}, \texttt{seaborn} \citep[][]{Waskom:2021}, \texttt{specutils} \citep[][]{Specutils:2019}}

\bibliographystyle{aasjournal}
\bibliography{main}{}
\allauthors

\end{document}